\documentclass[structabstract]{aa}
\usepackage{graphicx}
\usepackage{txfonts}
\usepackage{longtable}
\begin{document}
   \title{The outskirts of Cygnus OB2
\thanks{Based on observations collected at the Centro Astron\'omico
Hispano-Alem\'an (CAHA) at Calar Alto, operated jointly by the
Max-Planck Institut f\"ur Astronomie and the Instituto de Astrof\'\i
sica de Andaluc\'\i a (CSIC)}
   }

%\subtitle{I. Overviewing the $\kappa$-mechanism}

   \author{F. Comer\'on\inst{1}\thanks{Visiting astronomer at the
   Vatican Observatory}
             \and
           A. Pasquali\inst{2}
             \and
           F. Figueras\inst{3}
             \and
           J. Torra\inst{3}
          }

   \institute{European Southern Observatory, Karl-Schwarzschild-Strasse 2,
              D-85748 Garching, Germany\\
          \email{fcomeron@eso.org}
              \and
              Max-Planck-Institut f\"ur Astronomie, K\"onigstuhl 17, D-69117 Heidelberg, Germany\\
          \email{pasquali@mpia-hd.mpg.de}
              \and
              Departament d'Astronomia i Meteorologia, Universitat de
              Barcelona, E-08028 Barcelona, Spain\\
          \email{jordi@am.ub.es, cesca@am.ub.es}
             }

   \date{Received; accepted}

% \abstract{}{}{}{}{}
% 5 {} token are mandatory

  \abstract
   {Cygnus~OB2 is one of the richest OB associations in the local Galaxy, and is located
in a vast complex containing several other associations, clusters,
molecular clouds, and HII regions. However, the stellar content of
Cygnus~OB2 and its surroundings remains rather poorly known largely
due to the considerable reddening in its direction at visible
wavelength.}
   {We investigate the possible existence of an extended halo of early-type stars around
Cygnus~OB2, which is hinted at by near-infrared color-color
diagrams, and its relationship to Cygnus~OB2 itself, as well as to
the nearby association Cygnus~OB9 and to the star forming regions in
the Cygnus~X North complex.}
   {Candidate selection is made with photometry in the 2MASS all-sky point source catalog.
The early-type nature of the selected candidates is confirmed or
discarded through our infrared spectroscopy at low resolution. In
addition, spectral classifications in the visible are presented for
many lightly-reddened stars.}
   {A total of 96 early-type stars are identified in the targeted region,
which amounts to nearly half of the observed sample. Most of them
have featureless near-infrared spectra as expected from OB stars at
the available resolution. Another 18 stars that display Brackett
emission lines can be divided between evolved massive stars (most
likely Be stars) and Herbig Ae/Be stars based on their infrared
excesses. A component associated with Cygnus~OB9/NGC~6910 is clearly
identified, as well as an enhancement in the surface density of
early-type stars at Cygnus~X North. We also find a field population,
consisting largely of early B giants and supergiants, which is
probably the same as identified in recent studies of the inner
$1^\circ$ circle around Cygnus~OB2. The age and large extension of
this population discards a direct relationship with Cygnus~OB2 or
any other particular association.}
   {Earlier claims of the possible large extent of Cygnus~OB2 beyond
its central, very massive aggregate seem to be dismissed by our
findings. The existence of a nearly ubiquitous population of evolved
stars with massive precursors suggests a massive star formation
history in Cygnus having started long before the formation of the
currently observed OB associations in the region.}


   \maketitle
%
%________________________________________________________________

\section{Introduction}

  \object{Cygnus~OB2} is one of the richest OB associations known in our
Galaxy, and it is the most nearby one of its kind at less than 2~kpc
from the Sun (Kn\"odlseder~\cite{knoedlseder03}). It is also a
unique resource for the observational study of the upper end of the
stellar mass function (e.g., Herrero et al.~\cite{herrero99}): O
stars, Wolf-Rayet (WR) stars, Luminous Blue Variables (LBVs), B[e]
stars... are all classes with representatives in Cygnus~OB2. This
association thus provides, in a single complex, the most complete
and accessible showcase of the variety found among the hottest and
most massive stars.

  $UBV$ photometry reveals the presence of hundreds of early-type,
heavily-obscured stars in Cygnus~OB2 (Massey \&
Thompson~\cite{massey91}; Kiminki et al.~\cite{kiminki07}). However,
a more real measure of its richness is given by star counts in the
near infrared, where the large, patchy foreground extinction in the
direction of the association is greatly reduced. Using this
technique, Kn\"odlseder~(\cite{knoedlseder00}) has inferred a
content of $\sim 2600$~OB stars, including over 100 O-type stars and
evolved stars having massive progenitors. Support for this estimate
has been provided by Comer\'on et al.~(\cite{comeron02}) using
near-infrared low resolution spectroscopy. High signal-to-noise
ratio spectra in the visible of the least obscured objects of this
sample carried out by Hanson~(\cite{hanson03}) has confirmed the
classification of the vast majority of the early-type star
candidates listed by Comer\'on et al.~(\cite{comeron02}). However,
it has also shown that a large fraction of them are giant and
supergiant B stars rather than O stars. This suggests a picture more
complex than that of a single, recent starburst less than 3~Myr ago
giving rise to the association (Massey \& Thompson~\cite{massey91}).
It also casts doubts on the actual membership of these stars in
Cygnus~OB2, given their older ages and their extended spatial
distribution when compared to the strongly-clustered, earliest-type
stars at the center of the association. Since the precursors of the
extended population dominated by early B-type giants and supergiants
must have been O-type stars, the region must have been producing
massive stars long before the formation of the major young clusters
and associations currently observed there.

  We investigate the nature of this population and its
possible relationship with Cygnus~OB2 by studying the early-type
stellar content in a region lying between $1^\circ$ an $2^\circ$
from the center of the association. Earlier results of this work
have already revealed the existence of interesting objects possibly
related to Cygnus~OB2 well beyond the boundaries of the central
cluster. Examples are \object{WR~142a}, a WC8 Wolf-Rayet star about
$1.3^\circ$ from the center of Cygnus~OB2 (Pasquali et
al.~\cite{pasquali02}), and the very massive O4If runaway star
\object{BD$+43^\circ 3654$} more than two degrees away (Comer\'on \&
Pasquali~\cite{comeron07}). The study of the outskirts of Cygnus~OB2
is also motivated by its overlap with other structures tracing
recent star formation, most notably the Cygnus~OB9 association and
several compact HII regions and embedded clusters belonging to the
Cygnus~X molecular cloud complex (Schneider et
al.~\cite{schneider06}).

  We use the same technique based on 2MASS $JHK$
colors used in Comer\'on et al.~(\cite{comeron02}) to identify OB
stars in the targeted region. Infrared spectroscopy allows us to
identify new young intermediate-mass stars and evolved massive
stars, both characterized by their emission in Br$\gamma$, as well
as other likely early-type stars without noticeable emission lines.
We also provide spectral types of many of these objects whose
relatively light obscuration makes them accessible to spectral
classification in the blue. This paper can thus be regarded as a
continuation of the previous work of Comer\'on et
al.~(\cite{comeron02}) and Hanson~(\cite{hanson03}), now probing the
largely uncharted region beyond the limits of Cygnus~OB2 explored
thus far. The reader is referred to the recent review by Reipurth \&
Schneider~(\cite{reipurth08}) for a detailed description of the
current state of knowledge of the vast extent of Cygnus star forming
complexes, and, in particular, of the Cygnus~OB2 association and the
Cygnus~X region.

\section{Target selection~\label{target_selection}}

  The use of the $(J-H)$, $(H-K_S)$ diagram to distinguish distant
early-type stars from late-type evolved stars is founded on the
difference in intrinsic colors between both populations. The
interstellar reddening law keeps the {\it loci} occupied by bright,
reddened early- and late-type stars separated by a gap of $\sim
0.4$~mag in $(J-H)$ at any given $(H-K_S)$ color, in the sense of
the $(J-H)$ color being bluer; see, e.g., Comer\'on \&
Pasquali~(\cite{comeron05}) for an application of this technique to
the identification of the star responsible for the ionization of the
North America and Pelican nebulae. The location of a star along the
band defined by the reddened colors of early-type stars obscured by
different amounts is not a proof of its early spectral type, since a
certain degree of contamination of this {\it locus} is caused by
evolved stars with particular photometric properties. Nevertheless,
the signature of a rich association such as Cygnus~OB2 in the
$(J-H)$, $(H-K_S)$ diagram is clearly indicated by the presence of
an enhancement of the population of this {\it locus} (see, e.g.,
Fig.~2 of Comer\'on et al.~\cite{comeron02}). This makes the
early-type population easily identifiable even in regions where its
areal density is far below that of cool giants, thus extending the
sensitivity threshold to the early-type population of the
association well beyond the $\simeq 1^\circ$ radius where it stands
out in infrared starcounts.

\begin{figure}
 \centering
 \includegraphics[width=8.5cm]{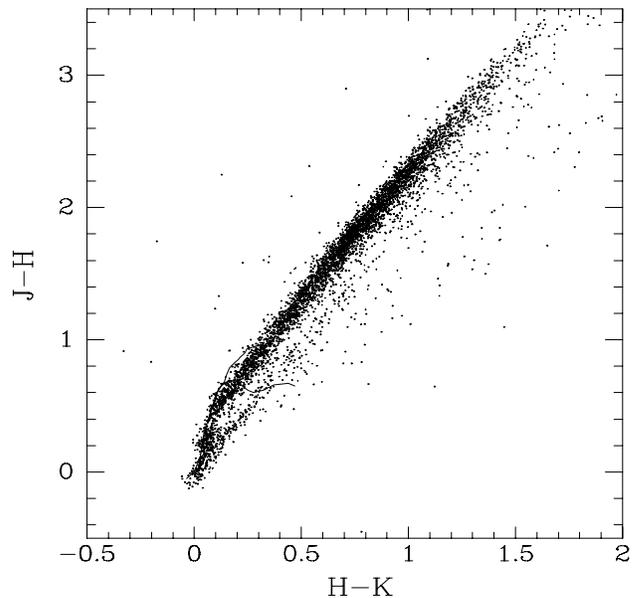}
 \caption[]{$(J-H), (H-K)$ diagram of the field surrounding
Cygnus~OB2 between $1^\circ$ and $2^\circ$ from the approximate
center of the association, at $\alpha(2000) = 20^h 32^m 26^s$,
$\delta(2000) = +40^\circ 52'$. The plotted magnitudes are from the
2MASS catalog. Only objects brighter than $K_S = 10$ are included.
Candidate early-type stars are located along a band that has its
origin near $(J-H) = 0$, $(H-K) = 0$ and traces the reddening
vector.}
  \label{fig_colcol2mass}
\end{figure}

  The presence of early-type stars beyond $1^\circ$ from the
center of Cygnus~OB2 is clearly indicated in
Fig.~\ref{fig_colcol2mass}, where we plot the 2MASS $(J-H)$,
$(H-K_S)$ diagram of point sources located between $1^\circ$ and
$2^\circ$ from the approximate center of Cygnus~OB2 at $\alpha(2000)
= 20^h 32^m 26^s$, $\delta(2000) = +40^\circ 52'$. A band of objects
with varying reddening runs along the {\it locus} of early-type
stars, well separated from the dominant band defined by cool
background stars, which runs above it. It is interesting to note
that the number of candidate OB stars in the region between
$1^\circ$ and $2^\circ$ from the center of Cygnus OB2 is comparable
to that found within $1^\circ$ from the center. In addition, the
spread in extinctions is larger than in the central region,
indicating that many OB stars may be very obscured and thus
inaccessible to survey in the visible.

  To confirm the nature of individual stars lying in the early-type
{\it locus} described above and their likely membership in the
association, we obtained low-resolution near-infrared spectroscopy
of 232 stars in the region between $1^\circ$ and $2^\circ$ from the
center of the association. The selected stars have $(J-H)-1.70
(H-K_S) < 0.20$, implying a closer position to the {\it locus} of
early-type stars than to that of late-type stars; and $K_S < 8.8$,
allowing us to obtain, with a reasonable investment of observing
time, a large number of $K$-band spectra. We also obtained
classification-quality spectra in the blue of 30 of the stars with
near-infrared spectra consistent with early types and membership in
the association, and whose foreground obscuration was relatively low
as judged from their $B$ magnitude listed in the USNO-B catalog
(Monet et al.~\cite{monet03}).

  We adopt here the rather loose denomination 'Cygnus~OB2 halo members'
to refer to early-type stars and young stellar objects located in
the area under discussion, and suspected to lie at the same distance
as the association. This is not meant to indicate confirmed
membership in the Cygnus~OB2 association. Moreover, some of the
stars we discuss are actually located within the boundaries of
\object{Cygnus~OB9} (including some members of the open cluster
\object{NGC~6910} belonging to that association) or are associated
with star forming regions in \object{Cygnus~X} like \object{DR~17}
or \object{DR~21} that are normally not considered to be a part of
Cygnus~OB2.

  Hanson~(\cite{hanson03}) has produced a detailed discussion on the
distance to Cygnus~OB2 based on current available calibrations of
the intrinsic properties of upper main sequence stars. Following
these conclusions, we adopt a distance of 1.45~kpc (distance modulus
10.80) to the region. We also assume that all the structures
discussed here belong to a physically-coherent region lying at
approximately that distance, rather than belonging to disconnected
entities along the line of sight, as further discussed in
Sect.~\ref{distribution}.

\section{Observations\label{observations}}

\subsection{Near-infrared spectroscopy \label{spectroscopy_ir}}

  We carried out near-infrared spectroscopic observations during two
observing runs at the Calar Alto observatory. The first one took
place from 16 June to 1 July 2002 at the 1.23m telescope, and the
second one from 31 July to 2 August 2004 at the 2.2m telescope. The
instrument used was in both cases MAGIC, a NICMOS3-based
near-infrared camera and spectrograph. We obtained the spectra using
the resin-replica grism providing a resolution $\lambda / \Delta
\lambda = 240$ over the 1.50-2.40~$\mu$m range with the $1''$ slit
used. Each star was observed at six positions along the slit, with
exposure times per position ranging from 20~s (stacking 10
integrations of 2~s) for the brightest stars to 60~s (stacking 20
integrations of 3~s) for the faintest, with exposure times
determined by the 2MASS $K_S$ magnitude and the aperture of the
telescope used. We carried out the extraction and calibration of the
spectra with dedicated IRAF\footnote{IRAF is distributed by NOAO,
which is operated by the Association of Universities for Research in
Astronomy, Inc., under contract to the National Science Foundation.}
scripts. The frames obtained at consecutive slit positions were
subtracted from each other to cancel out the sky contribution to the
spectrum, and were divided by a flat field frame.  A one-dimensional
object spectrum was then extracted from each sky-subtracted,
flat-fielded frame. We performed the wavelength calibration of each
individual spectrum with the OH airglow lines in each frame as a
reference (Oliva \& Origlia~\cite{oliva92}). The
wavelength-calibrated spectra extracted at each sky position were
then coadded, with deviant pixels due to detector defects or cosmic
ray hits automatically clipped off. Cancellation of telluric
features was achieved by ratioing the object spectra by those of the
nearby G5IV star HD190771, which is expected to be featureless at
the resolution employed, reduced in the same manner. Finally, we
performed relative flux calibration by multiplying the reduced
spectra by that of a 5,700~K blackbody, which should be a good
approximation to the spectral energy distribution of an unreddened
G5IV star in the wavelength range covered by our spectra.

\subsection{Visible spectroscopy \label{spectroscopy_vis}}

  Spectroscopy in the visible of 26 targets suspected to be reddened
early-type stars from their near-infrared photometric and
spectroscopic properties was obtained at Calar Alto using CAFOS, the
facility visible-light imager and low-resolution spectrograph, in an
observing run between 1 and 12 July 2005. We obtained spectra of
four additional stars with an identical instrumental setup at the
same telescope in an observing run between 17 and 25 August 2006.
The grism used covered the range shortwards of $\lambda = 6350$~\AA\
at a resolution of $\lambda / \Delta \lambda = 1000$ and with a
$1''5$ slit. We based the exposure times on the $B$ magnitude listed
in the USNO-B catalog, and ranged from 10 min to 180 min. For
integration times longer than 30~min, the exposures were split in
blocks of 30 min, and the individual extracted and reduced spectra
were then stacked together. Spectra of three lamps of HgCd, He, and
Rb were taken between subsequent exposures for wavelength
calibration, to minimize the effects of instrument flexure. The
frames containing the raw spectra were subtracted from bias and
divided by a spectroscopic flat field, and the spectra were
subsequently extracted from each one of them. The individual
wavelength-calibrated spectra were coadded after identification and
removal of cosmic ray hits. The coadded spectra were then normalized
to the interpolated continuum, to facilitate feature recognition and
comparison to spectral atlases of these generally heavily-reddened
objects.

\section{Results\label{results}}

\subsection{Infrared spectral classification and membership\label{class_ir}}

  Despite the limited resolution and signal-to-noise ratio of our infrared
spectra, it is possible to broadly classify our entire sample of
stars into eight classes:

\begin{itemize}

\item {\it Reddened stars with featureless spectra:} the absence
of spectral features at the resolution and signal-to-noise ratio of
our spectra is expected from hot stars earlier than mid-B type
(Hanson et al.~\cite{hanson96}), at which the Brackett lines begin
to be clearly visible. Their significant foreground reddening and
the relatively bright limiting magnitude of our sample ensures that
these are luminous early-type stars. Whereas G-type stars would also
appear featureless in our spectra, their infrared colors would in
general exclude them from our sample (Sect.~\ref{target_selection}),
and main sequence G stars would be too faint at the distances where
reddening begins being noticeable. The sample of reddened stars with
featureless spectra (see examples in Fig.~\ref{fig_featureless}) is
thus expected to have little or no contamination by objects other
than O and B stars, as confirmed by our visible spectroscopy of some
of the least heavily-reddened stars in this category. A similar
conclusion was reached by Hanson~(\cite{hanson03}), who obtained
visible spectroscopy of 14 out of the 31 stars selected in the same
manner by Comer\'on et al.~(\cite{comeron02}), and confirmed that
all of them are indeed early-type stars. Given the rarity of such
objects in the field, we consider that this class provides a largely
uncontaminated sample of members of the Cygnus region. This is the
class with most members in our sample, and the 66 stars belonging to
it are listed in Table~\ref{tab_red_featureless}\footnote{Tables 1
to 9 are only available in the online version of this paper.}.
Visible spectral classifications exist for 28 of these stars, of
which 24 are presented in this work for the first time.

\begin{figure}
 \centering
 \includegraphics[width=7cm]{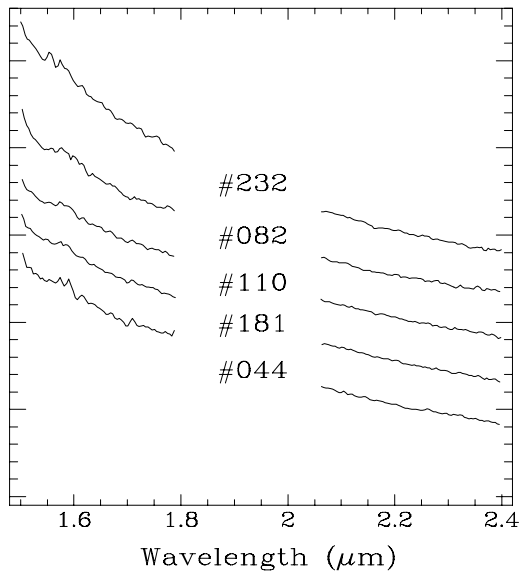}
 \caption[]{Typical examples of suspected reddened, early-type
stars. These stars appear featureless at the resolution and
signal-to-noise ratio of our near-infrared spectra.}
  \label{fig_featureless}
\end{figure}

\item {\it Reddened stars without measurable
absorption lines, but with emission lines (particularly
Br$\gamma$):} stars in this category are expected to belong to two
distinct, unrelated classes. The first one is composed of evolved
massive stars such as B[e], Of, LBV stars or, more likely, early Be
stars (Morris et al.~\cite{morris96}), all of which have massive
precursors. The second class is composed of intermediate-mass Herbig
Ae/Be stars. We consider both classes of objects as possible members
of the Cygnus complex as well, and we discuss in Sect.~\ref{colcol}
how we can distinguish between both. We list in
Table~\ref{tab_emission} the 19 stars in this category, whose
spectra are plotted in Fig.~\ref{fig_emission}. Visible spectra are
available for 5 of them, and a sixth one, the Wolf-Rayet star
WR~142a, has been classified as WC8 by Pasquali et
al.~(\cite{pasquali02}) from its infrared spectrum. Six of those
stars, including WR~142a, had been previously recognized in
objective prism surveys as emission-line stars.

\begin{figure}
 \centering
 \includegraphics[width=7cm]{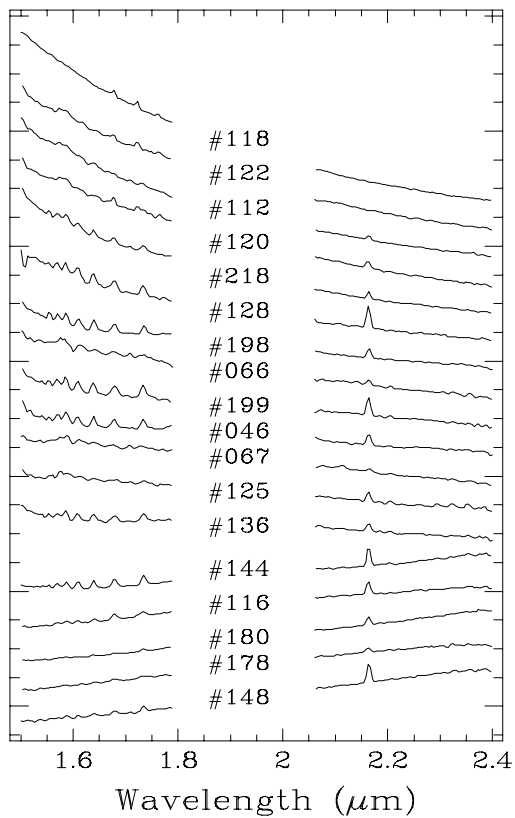}
 \caption[]{Near-infrared spectra of the stars in our sample that display
Brackett-series lines in emission. For convenience, the spectra are
sorted by increasing reddening from top to bottom. The spectrum of
an additional emission-line star, the Wolf-Rayet WR~142a, can be
found in Pasquali et al.~(\cite{pasquali02}). The sample shown here
is expected to contain representatives of two different classes,
classical Be stars and intermediate-mass Herbig Ae/Be stars, with
the reddest stars belonging to the latter category.}
  \label{fig_emission}
\end{figure}

\item {\it Lightly or no reddened stars with featureless spectra:} A
total of 31 stars in our sample show featureless infrared spectra,
but their colors indicate little reddening in their direction, thus
casting doubts as to their membership in the Cygnus complex.
Fortunately, spectral classifications in the visible, as well as
photometric classifications based on Vilnius photometry, are
available for most of these stars, showing that the sample can be
split into two groups:

\begin{itemize}

\item O and early B stars with light reddening, whose apparent
magnitudes are consistent with the adopted distance modulus of
Cygnus~OB2. We consider these stars as likely members of the Cygnus
complex, placed in the least reddened parts of the region, and we
list them in Table~\ref{tab_lightlyred_members}. We have included in
it star \#30, classified as B8 on the basis of its Vilnius
photometry. Such spectral type is unlikely, however, as a B8 star
should show noticeable Brackett-series lines in our near-infrared
spectra. An earlier B-type seems more likely, and we thus consider
it as a possible member of the complex, although with an uncertain
spectral type.

\item Virtually unreddened F-G stars, some of which have high proper
motions. We consider those as non-members and we list them in
Table~\ref{tab_foreground_oth}.

\end{itemize}

  Only 11 stars in this class, mostly faint, have not been observed
spectroscopically and it is not possible to assign them to either of
those two subclasses. To preserve our sample of Cygnus complex
members as uncontaminated as possible, we list these stars
separately within Table~\ref{tab_foreground_oth} and we do not
consider them among the possible members.

\begin{figure*}
 \includegraphics[width=12cm]{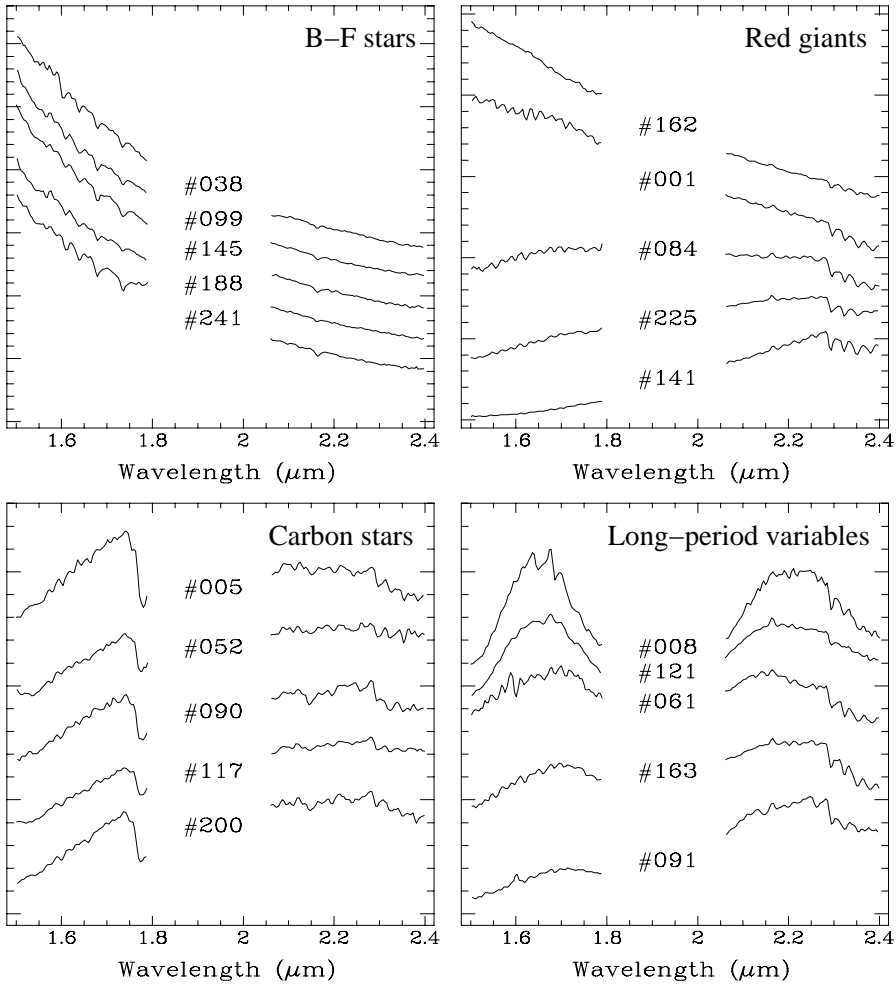}
 \caption[]{Typical near-infrared spectra of the different classes of
stars contaminating our sample of color-selected, early-type
candidate members of the Cygnus~OB2 region, which are described in
Sect.~\ref{class_ir}.}
  \label{fig_contam}
\end{figure*}

\item {\it Stars with obvious Br$\gamma$ absorption, and possibly
other Brackett lines as well:} these are unambiguously classified as
late-B, A or early-F stars. Their infrared colors generally indicate
little or no foreground extinction, suggesting distances closer than
those of the members of Cygnus~OB2. With very few exceptions,
spectral classifications in the visible exist in the literature for
most of these stars, confirming the above range of spectral types.
We thus consider the 48 lightly-reddened objects that we identify in
this category, listed in Table~\ref{tab_foreground_BF}, as
foreground stars. A few exceptions are found: 4 stars have clearly
reddened colors and still display obvious Brackett absorption lines
indicating spectral types in the late-B to early-F range. Since
their apparent magnitudes and reddening indicate an intrinsically
high luminosity, these objects are probably giants or supergiants.
Their possible membership in the Cygnus~region is unclear, and we
will not include them in our census of possible members. We list
them separately in Table~\ref{tab_foreground_BF} and display some
typical spectra in Fig.~\ref{fig_contam}.

\begin{figure}
 \centering
 \includegraphics[width=7cm]{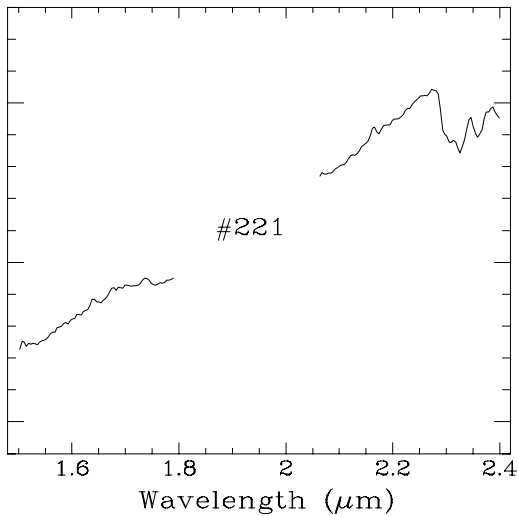}
 \caption[]{Spectrum of the likely young stellar object \#221 in the
DR~21 star forming region.}
  \label{fig_221}
\end{figure}

\item {\it Red giant branch stars:} Although the infrared colors of
these stars normally place them on the {\it locus} of reddened
late-type stars, infrared excesses at 2~$\mu$m can move them to the
region of interest of the $(J-H)$, $(H-K_S)$ diagram. This seems to
be the case of most of the 16 stars in our sample that clearly
display CO bands starting at 2.29~$\mu$m despite having been
color-selected. They are listed in Table~\ref{tab_rgb}, and display
a wide range of foreground reddening, as is obvious in the
representative examples shown in Fig.~\ref{fig_contam}. We note,
however, that one member of this class, star \#221, is projected on
the DR~21 star forming region, appears extended at 2.2~$\mu$m and
displays a spectral energy distribution at longer wavelengths (Davis
et al.~\cite{davis07}) characteristic of the Class~I/Class~II
transition (Adams et al.~\cite{adams87}). It is thus likely to be a
young stellar object belonging to DR~21, and still surrounded by
substantial amounts of gas and dust with its CO absorption being
produced in its circumstellar envelope. At the available resolution
and signal-to-noise ratio its spectrum (Fig.~\ref{fig_221}) is
typical of a cold photosphere, except perhaps for its broader CO
bands.

\item {\it Carbon stars:} Their frequent infrared excesses also
place carbon stars in the region of the $(J-H)$, $(H-K_S)$ diagram
occupied by early-type stars. They are easily recognized by the
numerous features in the $K$ window, in addition to the CO bands,
and, most particularly, by the prominent C$_2$ band at 1.77~$\mu$m
just below the telluric H$_2$O cutoff (Lan\c{c}on \&
Wood~\cite{lancon00}; Fig.~\ref{fig_contam}). We find 29 such stars
in our sample, which are listed in Table~\ref{tab_carbon}. Only the
two brightest ones in $K_S$ had been previously recognized as carbon
stars, and most of the others are IRAS point sources. We also note
that one of the stars that we classify as a red giant in
Table~\ref{tab_rgb}, star \#9, has been identified as a carbon star
by Chen \& Chen~(\cite{chen03}), although we do not identify its
characteristic features in the near-infrared.

\item {\it Long-period variables:} the infrared colors of Mira and
semiregular long-period variables makes them another potential
contaminant of the early-type {\it locus}. The 14 members of this
class that we identify in our sample (Table~\ref{tab_mirae};
Fig.~\ref{fig_contam}) are easily distinguished by their prominent
CO bands as well as by the marked broad absorption due to H$_2$O in
their extended envelopes that dominates the spectral energy
distribution in the region above and below the telluric feature
separating the $H$ and $K$ bands.

\begin{figure}
 \centering
 \includegraphics[width=7cm]{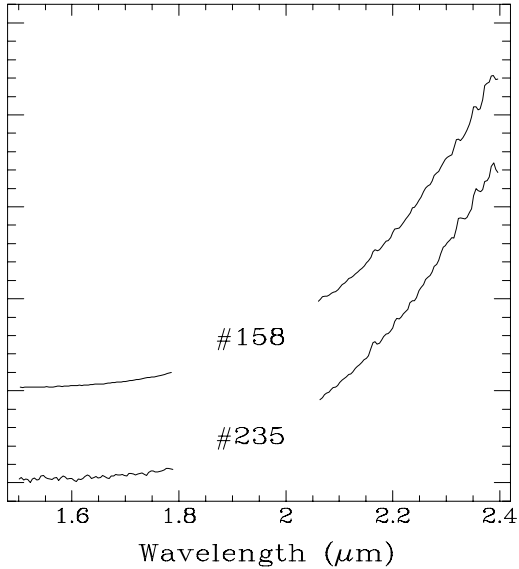}
 \caption[]{Spectra of the extremely red sources \#158 and \#235.
Star \#158 is the well-studied massive protostellar object AFGL~2591
(see text). Note the similar appearance of the weak Br$\gamma$
emission line at 2.166~$\mu$m.}
  \label{fig_protostars}
\end{figure}

\item {\it Protostellar sources:} two objects in our sample display
extremely red colors ($H-K_S = 2.76$ and $4.28$, respectively) and
featureless spectra, except for hints of Br$\gamma$ emission
(Table~\ref{tab_extremered}). The first one is identified as the
well-known source \object{AFGL~2591}, a high-mass protostellar
object that has been the subject of numerous studies; see, e.g.,
Benz et al.~(\cite{benz07}), Poelman \& van der
Tak~(\cite{poelman07}), as well as the review by Reipurth \&
Schneider~(\cite{reipurth08}) for recent discussions on this source.
The second one is identified as \object{IRAS~20249+3953}, for which
no detailed observations have been published thus far. The obvious
similarity of its near-infrared spectrum to that of AFGL~2591, as
seen in Fig.~\ref{fig_protostars}, leads us to classify it as a
massive protostar also, perhaps in an earlier evolutionary stage
than AFGL 2591, given its much redder colors. The source
IRAS~20249+3953 was undetected in the search for SiO maser emission
by Nakashima \& Deguchi~(\cite{nakashima03}).

\end{itemize}

\onltab{2}{
\begin{table*}
\caption[]{Emission line stars\label{tab_emission}}
\begin{tabular}{cccccccll}
\hline
Star & $\alpha (2000)$ & $\delta(2000)$ & $J$ & $H$ & $K_S$ & Vis. classification & Other IDs & Notes \\
\hline
  46 & 20  24  38.42 & 41  14  09.9 &   8.574 &  7.762 &  7.055 & & \object{AS~415} & 1 \\
  48 & 20  24  06.20 & 41  25  33.9 &   9.220 &  8.077 &  7.089 & & \object{WR142a} & 2 \\
  66 & 20  26  52.54 & 41  58  38.2 &   8.703 &  7.880 &  7.340 & & & 3 \\
  67 & 20  36  44.24 & 41  53  17.4 &   9.077 &  7.986 &  7.348 & & & \\
 112 & 20  28  15.21 & 42  25  39.1 &   8.286 &  7.999 &  7.838 & B2V & \object{BD+41~3762} & \\
 116 & 20  38  08.27 & 42  38  36.5 &  12.058 &  9.687 &  7.874 & & & 4, 5 \\
 118 & 20  35  16.63 & 42  50  47.1 &   8.239 &  8.018 &  7.906 & & \object{BD+42~3793} & 6 \\
 120 & 20  38  59.18 & 42  02  39.4 &   8.686 &  8.191 &  7.909 & B0Ib & \object{AS~434} & 7 \\
 122 & 20  27  25.52 & 39  29  24.5 &   8.499 &  8.120 &  7.923 & B0.2IV & & 8 \\
 125 & 20  38  13.55 & 41  44  16.3 &   9.865 &  8.635 &  7.952 & & & \\
 128 & 20  28  39.20 & 42  46  28.1 &   8.883 &  8.392 &  7.997 & B0IIIe & \object{AS~419} & 9 \\
 136 & 20  23  16.63 & 41  09  25.0 &  10.106 &  8.937 &  8.120 & & & 5 \\
 144 & 20  38  45.88 & 42  07  04.8 &  10.666 &  9.603 &  8.165 & & & 5 \\
 148 & 20  38  57.19 & 42  22  40.9 &  13.160 & 10.307 &  8.214 & & & 5, 10 \\
 178 & 20  30  41.44 & 39  45  15.5 &  13.083 & 10.379 &  8.464 & & & 5, 11 \\
 180 & 20  25  28.57 & 41  08  26.1 &  12.325 & 10.207 &  8.464 & & & 5 \\
 198 & 20  25  59.59 & 40  10  18.4 &   9.735 &  9.067 &  8.586 & & & 12 \\
 199 & 20  38  30.39 & 42  28  26.4 &  10.198 &  9.291 &  8.599 & & & 5, 13 \\
 218 & 20  38  12.88 & 40  57  16.9 &   9.222 &  8.894 &  8.699 & B2V & & \\
\hline\smallskip
\end{tabular}\\
{\it Notes:}
\smallskip\\
1: Identified as a star with very strong H$\alpha$ emission by
Merrill \& Burwell~(\cite{merrill50}).\\
2: Wolf-Rayet star, infrared spectral type WC8; see Pasquali et
al.~(\cite{pasquali02}). It had been previously recognized as an
emission-line star by Melikian \& Shevchenko~(\cite{melikian90}).\\
3: Already recognized as an emission line star by
Dolidze~(\cite{dolidze75}) (star 20-091).\\
4: In the W75N star forming region, strong excess at 8~$\mu$m (Davis
et al.~\cite{davis07}).\\
5: Suspected Herbig Ae/Be star basedon its infrared excess.\\
6: Double star, separation 1"1: type estimated as G3IV-V by
Straizys et al.~(\cite{straizys89}) based on Vilnius photometry.\\
7: Identified as a star with weak H$\alpha$ emission by Merrill \&
Burwell~(\cite{merrill50}).\\
8: In DR~6.\\
9: Identified as a star with strong H$\alpha$ emission by Merrill \&
Burwell~(\cite{merrill50}).\\
10: In the HII region \object{DR~21}.\\
11: Possibly associated to \object{IRAS~20288+3934}.\\
12: Already recognized as an emission line star by
Dolidze~(\cite{dolidze75}) (star 20-089). In the HII region \object{DR~5}.\\
13: In DR~21.\\
\end{table*}
}

\onltab{3}{
\begin{table*}
\caption[]{Lightly reddened OB stars\label{tab_lightlyred_members}}
\begin{tabular}{ccccccll}
\hline
Star & $\alpha (2000)$ & $\delta(2000)$ & $J$ & $H$ & $K_S$ & Other IDs & Notes \\
\hline
  27 & 20  25  22.12 &  40  13  01.2 &  6.836 & 6.623 & 6.495 & \object{HD~194649} & 1 \\
  30 & 20  38  10.94 &  42  10  26.0 &  7.047 & 6.670 & 6.556 & \object{BD+41~3833} & 2 \\
  31 & 20  26  20.91 &  39  40  10.3 &  7.105 & 6.802 & 6.580 & \object{V455~Cyg} & 3 \\
  50 & 20  25  55.09 &  41  20  11.8 &  7.167 & 7.147 & 7.116 & \object{HD~194779}& 4 \\
  64 & 20  28  30.24 &  42  00  35.3 &  7.334 & 7.351 & 7.318 & \object{HD~195229} & 5 \\
  68 & 20  23  10.46 &  40  45  52.4 &  7.727 & 7.473 & 7.353 & \object{BD+40~4146} & 6 \\
  71 & 20  27  36.77 &  42  02  07.6 &  7.271 & 7.327 & 7.384 & \object{HD~195089} & 7 \\
 142 & 20  22  44.75 &  40  42  52.8 &  8.056 & 8.129 & 8.152 & \object{HD~194194} & 8 \\
 150 & 20  27  43.62 &  40  35  43.6 &  8.415 & 8.316 & 8.250 & \object{BD+40 4179} & 9 \\
 238 & 20  38  28.89 &  40  09  56.8 &  8.888 & 8.848 & 8.810 & \object{BD+39~4263} & 10 \\
\hline\smallskip
\end{tabular}\\
{\it Notes:} \smallskip\\
1: Possibly associated to DR~5. Classified as O6.5 by Morgan et al.~(\cite{morgan55}).\\
2: Classified as B8 by Straizys et al.~(\cite{straizys89}) based on
Vilnius photometry.\\
3: Eclipsing binary (Giuricin et al.~\cite{giuricin83}), classified
as B2p?e? by Morgan et al.~(\cite{morgan55}).\\
4: Classified as B3II by Morgan et al.~(\cite{morgan55}).\\
5: Classified as B0.2III by Walborn~(\cite{walborn71}).\\
6: In cluster NGC~6910; classified as B1 by Walker \&
Hodge~(\cite{walker68}).\\
7: Classified as B2IV by Guetter~(\cite{guetter68}).\\
8: In cluster NGC~6910; classified as spectral type B1.5IV by
Walborn~(\cite{walborn71}).\\
9: Classified as O8V: by Morgan et al.~(\cite{morgan55}).\\
10: Ultraviolet-bright star, type B estimated by Carruthers \&
Page~(\cite{carruthers84}).\\
\end{table*}
}

\onltab{4}{
\begin{table*}
\caption[]{Foreground intermediate-type stars and lightly reddened,
unclassified stars\label{tab_foreground_oth}}
\begin{tabular}{ccccccll}
\hline
Star & $\alpha (2000)$ & $\delta(2000)$ & $J$ & $H$ & $K_S$ & Other IDs & Notes \\
\hline \noalign{\medskip {\it Foreground stars} \medskip}
  28 & 20  29  28.37 &  42  19  00.3 &  6.814 & 6.579 & 6.521 & \object{HD~195405} & 1 \\
  35 & 20  38  13.10 &  39  32  49.5 &  6.941 & 6.705 & 6.660 & \object{HD~196790}& 2\\
  57 & 20  24  36.12 &  41  30  02.8 &  7.603 & 7.281 & 7.192 & \object{LTT~15972} & 3 \\
  58 & 20  37  29.72 &  42  33  14.9 &  7.651 & 7.305 & 7.219 & \object{HIP~101763} & 4 \\
  69 & 20  25  39.03 &  41  53  45.6 &  7.673 & 7.428 & 7.370 & \object{HD~229293} & 5 \\
  73 & 20  24  03.81 &  40  20  22.7 &  7.653 & 7.454 & 7.391 & \object{HD~229244} & 6 \\
  85 & 20  29  46.12 &  39  39  01.1 &  7.842 & 7.572 & 7.516 & \object{BD+39~4212} & 7 \\
  92 & 20  39  34.11 &  42  01  30.3 &  7.784 & 7.670 & 7.616 & \object{BD+41~3844} & 8 \\
 174 & 20  29  52.91 &  42  21  27.4 &  8.714 & 8.481 & 8.423 & \object{BD+41 3776} & 9 \\
 203 & 20  40  07.90 &  41  15  10.8 &  9.067 & 8.708 & 8.615 & \object{BD+40~4272} & 10 \\
\noalign{\medskip {\it Lightly reddened stars with no classification
in the visible}
\medskip}
 139 & 20  35  35.00 &  40  01  29.5 &  8.380 & 8.211 & 8.128 & & \\
 191 & 20  38  46.80 &  42  01  06.4 &  8.908 & 8.615 & 8.552 & & \\
 197 & 20  26  36.09 &  40  28  29.1 &  8.810 & 8.606 & 8.583 & & \\
 201 & 20  24  22.65 &  41  26  08.5 &  8.901 & 8.652 & 8.612 & \object{HD~229266} & \\
 210 & 20  24  16.06 &  40  29  26.6 &  8.964 & 8.686 & 8.642 & \object{HD~229252} & \\
 214 & 20  39  28.12 &  40  36  15.1 &  9.020 & 8.747 & 8.677 & \object{BD+40~4266} & \\
 216 & 20  24  25.50 &  39  49  28.4 &  8.831 & 8.748 & 8.695 & \object{HD~229258} & \\
 231 & 20  30  02.17 &  42  14  34.1 &  9.053 & 8.845 & 8.767 & & \\
 234 & 20  23  59.70 &  40  07  46.8 &  8.952 & 8.820 & 8.783 & \object{HD~229237} & \\
 239 & 20  39  10.53 &  40  20  06.4 &  9.149 & 8.861 & 8.813 & & \\
 249 & 20  27  52.05 &  41  31  20.0 &  9.143 & 8.895 & 8.847 & & \\
\hline\smallskip
\end{tabular}\\
{\it Notes:} \smallskip\\
1: Classified as G0Iab: (Nassau \& Morgan~\cite{nassau52}) and as
G2IV (Bidelman~\cite{bidelman57}).\\
2: Classified as F8IV (Nordstr\o m et al.~\cite{nordstroem04}).\\
3: High proper motion, horizontal branch star; see e.g.
Behr~(\cite{behr03}).\\
4: Close binary, high proper motion star; Fabricius \&
Makarov~(\cite{fabricius00}).\\
5: Classified as K0 in the HD catalog.\\
6: Classified as G0 in the HD catalog.\\
7: Classified as F8 in the SAO catalog.\\
8: Classified as F0V by Straizys et al.~(\cite{straizys89}) based
on Vilnius photometry.\\
9: Classified as F5:V by Straizys et al.~(\cite{straizys89}) based
on Vilnius photometry.\\
10: Nearby, high proper motion star at 75~pc according to the
Hipparcos catalog. Suspected metal-poor subdwarf by Reid et al.~(\cite{reid01}).\\
\end{table*}
}

\onltab{5}{
\begin{table*}
\caption[]{Late B-F stars with Brackett absorption
lines\label{tab_foreground_BF}}
\begin{tabular}{cccccc}
\hline
Star & $\alpha (2000)$ & $\delta(2000)$ & $J$ & $H$ & $K_S$ \\
\hline \noalign{\medskip {\it Unreddened or lightly reddened stars}
\medskip}
  23 & 20  39  33.32  & 40  34  46.4  &  6.223 &   6.351 &  6.375 \\
  26 & 20  25  26.34  & 41  54  37.4  &  6.603 &   6.573 &  6.487 \\
  38 & 20  26   3.85  & 40  24   6.3  &  6.755 &   6.847 &  6.873 \\
  41 & 20  36  50.43  & 42  36  30.1  &  7.203 &    7.04 &  6.986 \\
  49 & 20  40   53.3  & 41  12    9.  &  7.012 &   7.082 &  7.097 \\
  53 & 20  25  51.05  & 39  37  53.0  &  7.540 &   7.231 &  7.153 \\
  59 & 20  26  43.77  & 39  29  45.4  &  7.154 &   7.214 &  7.219 \\
  77 & 20  35  12.48  & 42  51  15.3  &  7.765 &   7.595 &  7.457 \\
  78 & 20  31  26.34  & 39  56  19.7  &  7.481 &   7.481 &   7.46 \\
  83 & 20  34  33.06  & 41  59  47.9  &  7.532 &   7.542 &  7.499 \\
  95 & 20  26  51.15  & 41  51  33.5  &  7.889 &   7.682 &  7.651 \\
  99 & 20  25  28.57  & 39  47  36.7  &  7.625 &   7.703 &  7.698 \\
 104 & 20  28  42.56  & 39  28  25.1  &  7.915 &   7.744 &  7.743 \\
 106 & 20  26  46.23  & 39  27  44.1  &  7.719 &   7.749 &  7.778 \\
 107 & 20  25  15.68  & 40  59  55.1  &  7.912 &    7.82 &  7.783 \\
 126 & 20  26  43.76  & 40  20  50.1  &  7.848 &   7.932 &  7.967 \\
 145 & 20  23  21.83  & 41  30  37.7  &  8.246 &   8.212 &  8.192 \\
 147 & 20  36  24.25  & 39  11  40.7  &  8.207 &   8.191 &  8.213 \\
 149 & 20  29  23.57  & 40   5  56.5  &  8.164 &   8.181 &  8.225 \\
 153 & 20  27  30.84  & 41  26   1.2  &  8.329 &   8.276 &  8.265 \\
 154 & 20  31  59.74  & 42   3  38.3  &  8.276 &   8.286 &  8.276 \\
 155 & 20  38  34.33  & 41   9  37.4  &  8.469 &   8.28  & 8.279  \\
 157 & 20  32  44.44  & 39  13  15.2  &  8.508 &   8.338 &  8.301 \\
 161 & 20  42  17.28  & 41  29    4.  &  8.947 &   8.512 &  8.336 \\
 166 & 20  27  13.68  & 42  37  46.5  &  8.482 &   8.418 &  8.369 \\
 171 & 20  40  40.74  & 40  37  43.8  &  8.602 &   8.428 &  8.409 \\
 172 & 20  25  44.55  & 40   3  59.2  &   8.46 &   8.446 &   8.41 \\
 173 & 20  27   3.49  & 42  17  57.1  &  8.383 &   8.402 &  8.414 \\
 176 & 20  32   43.5  & 39  16  15.8  &  8.365 &   8.457 &  8.433 \\
 188 & 20  27  34.89  & 40  32  21.9  &  8.838 &   8.605 &   8.54 \\
 192 & 20  33  46.89  & 39  56  34.1  &  8.697 &    8.55 &  8.553 \\
 195 & 20  37  55.33  & 42   7  51.1  &  8.794 &   8.633 &  8.579 \\
 207 & 20  25  51.15  & 40  35  54.2  &  8.591 &    8.62 &  8.629 \\
 208 & 20  42   1.19  & 40  14  42.8  &  8.701 &   8.665 &  8.633 \\
 211 & 20  25   15.1  & 40   1   9.8  &  8.547 &    8.65 &  8.655 \\
 212 & 20  31  33.58  & 42  12  14.2  &  8.558 &   8.635 &   8.66 \\
 215 & 20  37  22.33  & 39  29   39.  &  8.688 &   8.679 &  8.691 \\
 217 & 20  28  10.76  & 39  12  47.8  &  8.746 &   8.758 &  8.696 \\
 223 & 20  37  34.28  & 39  48  27.8  &  9.088 &   8.847 &  8.723 \\
 226 & 20  41  59.45  & 41  26  58.9  &  8.865 &   8.787 &  8.737 \\
 228 & 20  36  34.73  & 41  48   8.9  &  8.923 &   8.779 &   8.74 \\
 233 & 20  24  58.78  & 39  50  21.7  &  8.713 &   8.776 &  8.776 \\
 236 & 20  39  38.95  & 40  35   11.  &  8.733 &   8.812 &  8.803 \\
 237 & 20  33  36.33  & 39  48  49.5  &  9.083 &   8.862 &  8.804 \\
 241 & 20  29  59.66  & 39  34  26.6  &  8.901 &   8.858 &  8.822 \\
 243 & 20  28  27.27  & 39  22  28.4  &  8.824 &   8.843 &  8.824 \\
 245 & 20  32  14.14  & 39  44   0.1  &  8.927 &    8.85 &  8.825 \\
 251 & 20  30  34.23  & 39  16  51.9  &  9.023 &   8.921 &  8.855 \\
\noalign{\medskip {\it Reddened stars}
\medskip}
 056 & 20 28 58.75  &  40 13 30.3 &  8.505 & 7.666 & 7.192 \\
 109 & 20 36 57.14  &  42 42 10.0 &  9.688 & 8.442 & 7.788 \\
 179 & 20 31 19.00  &  42  2 56.0 &  9.595 & 8.870 & 8.464 \\
 240 & 20 38 21.43  &  40 30 39.7 & 10.548 & 9.416 & 8.815 \\
\hline\smallskip
\end{tabular}
\end{table*}
}

\onltab{6}{
\begin{table}
\caption[]{Red giant candidates\label{tab_rgb}}
\begin{tabular}{ccccl}
\hline Star & $\alpha (2000)$ & $\delta(2000)$ & $K_S$ & Notes \\
\hline
  1 & 20  25  41.39 &  40  00  33.5 &  4.186 & 1 \\
  2 & 20  27  41.05 &  39  51  44.0 &  4.231 & \\
  3 & 20  24  17.99 &  41  13  09.9 &  4.309 & \\
  4 & 20  35  18.04 &  41  53  24.5 &  4.436 & 2 \\
  9 & 20  39  47.55 &  40  47  04.1 &  4.885 & 3 \\
 84 & 20  30  40.10 &  42  22  46.6 &  7.511 & \\
 87 & 20  36  40.68 &  40  07  29.4 &  7.553 & \\
127 & 20  31  00.20 &  42  20  16.1 &  7.976 & \\
141 & 20  39  22.70 &  41  13  11.4 &  8.145 & \\
162 & 20  33  53.54 &  42  58  14.8 &  8.337 & \\
168 & 20  22  39.72 &  40  31  40.4 &  8.388 & \\
182 & 20  35  28.30 &  42  53  00.6 &  8.471 & \\
184 & 20  24  22.85 &  39  58  23.6 &  8.473 & \\
221 & 20  38  46.36 &  42  24  39.7 &  8.713 & 4 \\
225 & 20  40  35.63 &  42  10  17.0 &  8.736 & \\
247 & 20  40  46.98 &  39  50  05.4 &  8.839 & \\
248 & 20  32  20.89 &  42  32  49.1 &  8.845 & \\
\hline\smallskip
\end{tabular}\\
{\it Notes:}\\
1: \object{HD~229291}, classified in the visible as M0
(Neckel~\cite{neckel58}).\\
2: \object{HD~196360}, classified in the visible as K0III
(Bidelman~\cite{bidelman57}).\\
3: Classified as a carbon star by Chen \& Chen~(\cite{chen03}); the
C$_2$ 1.77~$\mu$m feature is barely visible in our near-infrared spectrum.\\
4: Possible Class I/II member of DR~21 (Davis et
al.~\cite{davis07}).
\end{table}
}

\onltab{7}{
\begin{table}
\caption[]{Carbon stars\label{tab_carbon}}
\begin{tabular}{cccc}
\hline Star & $\alpha (2000)$ & $\delta(2000)$ & $K_S$ \\
\hline
  5 & 20  23  10.08 &  40  57  41.7 &  4.653 \\
  6 & 20  40  58.42 &  40  33  47.8 &  4.676 \\
 11 & 20  22  56.90 &  41  17  01.7 &  5.174 \\
 24 & 20  41  25.51 &  40  15  21.8 &  6.413 \\
 25 & 20  35  47.30 &  42  40  53.2 &  6.432 \\
 43 & 20  31  58.75 &  42  48  08.0 &  7.014 \\
 47 & 20  36  10.71 &  39  13  07.6 &  7.084 \\
 52 & 20  28  38.36 &  42  35  47.7 &  7.150 \\
 55 & 20  24  36.14 &  40  51  30.9 &  7.190 \\
 76 & 20  38  24.09 &  39  51  52.3 &  7.453 \\
 79 & 20  27  26.95 &  40  24  19.3 &  7.460 \\
 90 & 20  41  08.34 &  40  24  14.2 &  7.567 \\
 98 & 20  26  26.64 &  40  21  20.5 &  7.695 \\
102 & 20  33  11.10 &  39  02  24.3 &  7.731 \\
103 & 20  41  18.63 &  40  21  53.7 &  7.734 \\
105 & 20  29  14.98 &  39  19  23.5 &  7.750 \\
113 & 20  25  28.25 &  41  41  09.4 &  7.851 \\
114 & 20  30  50.87 &  42  38  14.4 &  7.852 \\
115 & 20  27  04.60 &  42  38  31.4 &  7.865 \\
117 & 20  23  16.58 &  41  07  44.0 &  7.876 \\
119 & 20  27  27.82 &  42  43  42.7 &  7.909 \\
123 & 20  28  24.74 &  42  20  10.2 &  7.924 \\
151 & 20  41  26.61 &  41  44  32.2 &  8.262 \\
165 & 20  34  45.66 &  42  49  13.3 &  8.359 \\
167 & 20  38  57.69 &  41  09  34.5 &  8.373 \\
187 & 20  38  18.30 &  41  42  39.5 &  8.529 \\
200 & 20  33  05.99 &  39  01  16.1 &  8.601 \\
213 & 20  29  36.10 &  41  49  24.9 &  8.674 \\
224 & 20  39  37.54 &  42  11  32.4 &  8.729 \\
\hline\smallskip
\end{tabular}
\end{table}
}

\onltab{8}{
\begin{table}
\caption[]{Likely long-period variables\label{tab_mirae}}
\begin{tabular}{ccccll}
\hline Star & $\alpha (2000)$ & $\delta(2000)$ & $K_S$ & Notes \\
\hline
  8 & 20  35  49.22 & 39  59  56.7 & 4.801 &  \\
 40 & 20  34  55.26 & 39  36  35.2 & 6.977 &  \\
 51 & 20  37  23.92 & 40  30  33.0 & 7.129 &  \\
 54 & 20  32  35.80 & 42  42  35.4 & 7.178 &  \\
 61 & 20  29  10.72 & 39  50  59.3 & 7.284 &  \\
 80 & 20  28  01.84 & 42  23  14.0 & 7.483 &  \\
 89 & 20  26  47.81 & 40  46  47.6 & 7.563 &  \\
 91 & 20  34  15.36 & 39  44  59.8 & 7.603 &  \\
111 & 20  27  23.51 & 39  42  01.6 & 7.832 &  \\
121 & 20  25  49.61 & 41  40  12.6 & 7.917 &  \\
124 & 20  39  10.34 & 39  33  49.0 & 7.931 &  \\
160 & 20  29  07.75 & 41  49  12.5 & 8.328 &  \\
163 & 20  27  46.56 & 41  32  02.4 & 8.345 &  \\
177 & 20  40  09.21 & 42  01  17.5 & 8.441 &  \\
\hline\smallskip
\end{tabular}
\end{table}
}

\onltab{9}{
\begin{table*}
\caption[]{Extremely red objects\label{tab_extremered}}
\begin{tabular}{ccccccll}
\hline
Star & $\alpha (2000)$ & $\delta(2000)$ & $J$ & $H$ & $K_S$ & IRAS source & Notes\\
\hline
158 & 20 29 24.36 &  40 11 15.6  &  12.325 & 11.082 & 8.320 & IRAS~20275+4001 & 1 \\
235 & 20 26 44.43 &  40 02 57.8  &  17.200 & 13.069 & 8.793 & IRAS~20249+3953 & 2 \\
\hline\smallskip
\end{tabular}\\
{\it Notes:}\\
1: Source AFGL~2591.\\
2: Possibly associated to DR~5.\\
\end{table*}
}

\subsection{Infrared color-color diagram\label{colcol}}

\begin{figure}
 \centering
 \includegraphics[width=9cm]{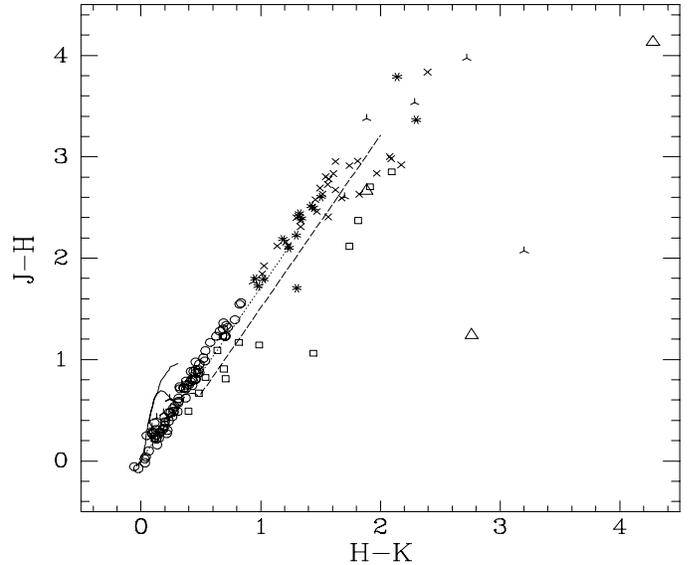}
 \caption[]{Color-color diagram of all the objects in our sample, with
the exception of foreground stars.{\it Circles:} stars with
featureless spectra; {\it squares:} emission-line stars; {\it
triangles:} massive young stellar objects; {\it 3-pointed
asterisks:} red giant branch stars; {\it 4-pointed asterisks:}
carbon stars; and {\it 8-pointed asterisks:} long-period variables.
The length of the reddening vector (dotted line) corresponds to
20~mag of visual extinction. The dashed line separates the {\it
loci} of classical Be from Herbig Ae/Be stars.}
  \label{fig_colcol}
\end{figure}

  The distribution of objects of our sample in the near-infrared
color-color diagram is presented in Fig.~\ref{fig_colcol}. The most
abundant class of stars, those displaying featureless near-infrared
spectra, are well aligned along the reddening vector, displaying the
colors of reddened normal early-type photospheres without
indications of near-infrared excess. Based on the position of the
reddest stars with the colors and near-infrared spectra of normal
photospheres, we estimate a maximum reddening toward the halo of
Cygnus~OB2 of $A_{V {\rm max}} \simeq 15$~mag, or $A_{K {\rm max}}
\simeq 1.7$~mag.

  The bluest emission-line stars tend to appear slightly to the
right of the limiting reddening vector having its origin at the
position of the earliest unreddened stars, thus indicating the
existence of moderate amounts of circumstellar excess. However, a
break is easily identified at $(H-K) >0.8$ as emission-line stars
with redder colors are all far more removed from the limiting
reddening vector. We interpret this as a consequence of the
existence of two classes of emission-line stars in our sample, as
discussed in Sect.~\ref{spectroscopy_ir}. Classical Be stars have
infrared excesses produced by free-free and free-bound transitions
in the ionized circumstellar gas, which is thought to be distributed
in a decretion disk formed as a consequence of mass loss from the
atmosphere of a star rotating at nearly the break-up speed (Sigut \&
Jones~\cite{sigut07}; Ekstr\"om et al.~\cite{ekstroem08}). Although
the Be phenomenon may appear at all stellar ages, and also among
young stars with a rapid initial rotation speed (Zorec \&
Briot~\cite{zorec97}), several observational lines of evidence
indicate that it most usually results from spin-up of the star via
mass transfer from a companion, near the end of the main sequence
lifetime (McSwain \& Gies~\cite{mcswain07}). On the other hand,
Herbig Ae/Be stars are pre-main sequence stars that display strong
infrared excesses produced by dust in their circumstellar accretion
disks, particularly in the puffed-up walls delimiting their inner
rims (Dullemond et al.~\cite{dullemond01}). As shown by Hern\'andez
et al.~(\cite{hernandez05}), the $JHK$ color-color diagram is a
useful tool for distinguishing between both classes of star and our
own sample confirms the split of both classes according to their
position in the $(J-H)$, $(H-K)$ diagram, or more precisely
according to the reddening-free parameter $Q = (J-H) - 1.70(H-K)$
already used in Sect.~\ref{target_selection}. Classical Be star
candidates cluster around $-0.18 < Q < 0.07$, whereas Herbig Ae/Be
candidates cover a broad range starting at $Q = -0.22$ and extending
to $Q=-1.38$. The adopted dividing line between classical Be and
Herbig Ae/Be stars, defined as $Q=-0.20$, is indicated in
Fig.~\ref{fig_colcol}. Since Herbig Ae/Be fulfill our selection
criteria due to their infrared excess rather than to their
photospheric colors, by including them in our sample of candidate
Cygnus~OB2 halo members we extend it to stars less massive than the
OB stars on which we focus. However, including the Herbig Ae/Be in
our discussion is still useful as they are additional tracers of
recent star formation.

\begin{figure*}
 \centering
 \includegraphics[width=18cm]{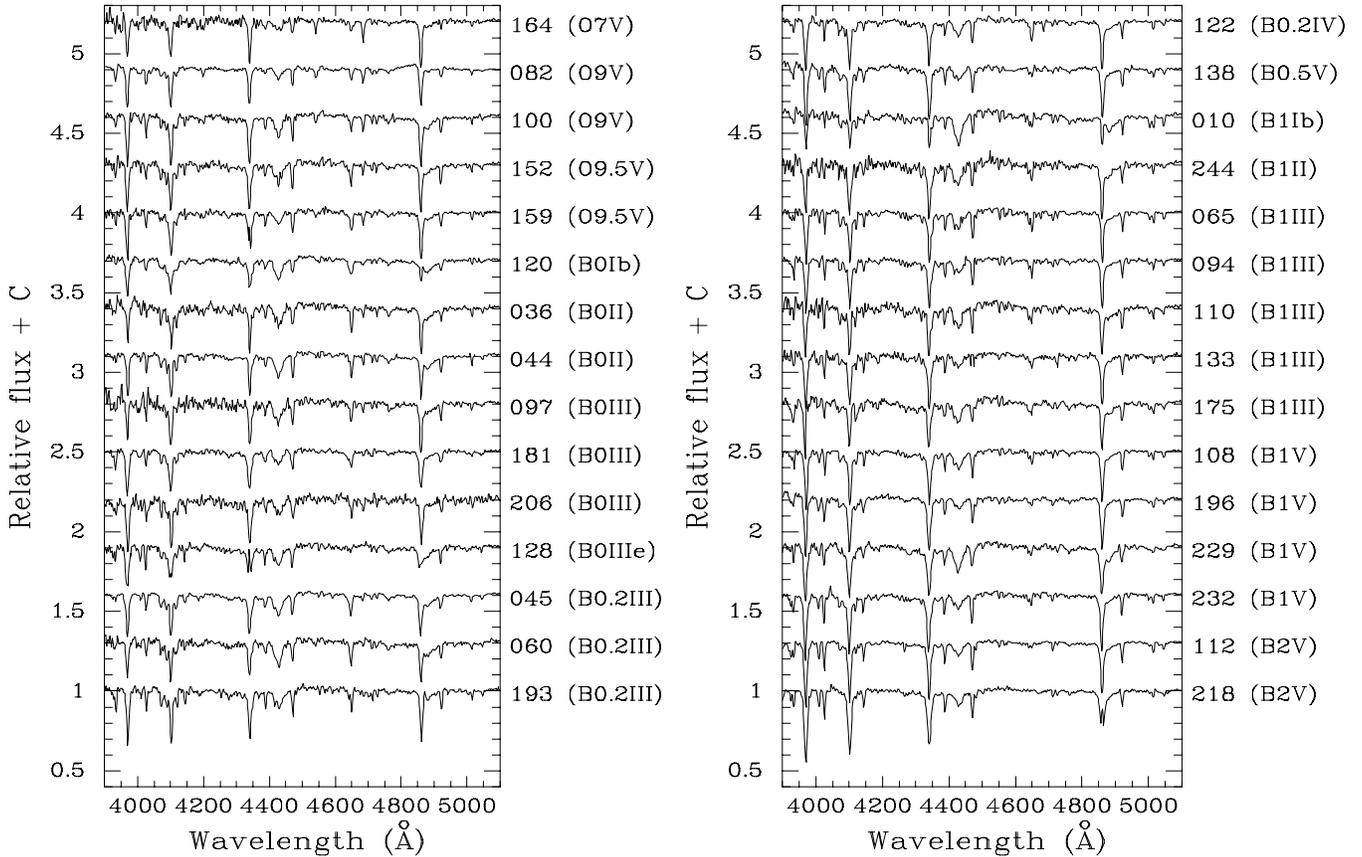}
 \caption[]{Visible spectra of the stars for which a new MK spectral
classification is given in this work. Note the strong interstellar
feature centered at 4428~\AA, due to the strong extinction. The main
criteria used for classification are outlined in
Sect.~\ref{spec_vis}.}
  \label{fig_specvis}
\end{figure*}

  Two of the massive young stellar objects that we include in our
sample stand out in Fig.~\ref{fig_colcol} due to their extreme
infrared excess. Star~\#158 (=AFGL~2591) is represented by the
triangle at $(J-H) = 1.24$, $(H-K) = 2.76$, whereas the even more
extreme star~\#235 (=IRAS~20249+3953) appears near the upper right
corner of the diagram. The candidate massive young stellar object
\#221 shows a modest amount of excess by comparison, occupying a
position similar to that of some Herbig Ae/Be stars, which probably
indicates a more evolved status. We note the apparent existence of a
third object with extreme infrared colors, star \#141, which we
classified among the red giant candidates on the basis of the
well-defined CO absorption bands longwards of 2.29~$\mu$m
characteristic of a late-type star. However, an inspection of the
2MASS images at the position of this star clearly shows the presence
of a nearby star of similar brightness at $K_S$, which dominates at
shorter wavelengths; in fact, star \#141 is not noticeable in the
2MASS $J$ image. The strong apparent infrared excess is thus almost
certainly an artifact due to contamination of the photometric
measurement by the dominating bright companion.

  The cool stars in our sample that we classify as non-members
appear almost entirely removed from the {\it loci} occupied by the
candidate members of the Cygnus~OB2 halo, having in general much
redder colors. This is partly caused by their intrinsically redder
spectral energy distributions, but the dominant factor is their
larger average distances due to their brighter infrared absolute
magnitudes. Whereas we expect OB stars in Cygnus~OB2 to have
absolute magnitudes in the $-3 < M_K < -5.5$ range (Martins \&
Plez~\cite{martins06}), cool red giants are expected to reach $M_K =
-7$ (e.g., Ferraro et al.~\cite{ferraro00}); long period variables
are in the range $-6.4 > M_K > -8.2$ (Knapp et al.~\cite{knapp03});
and carbon stars usually have $M_K < -8$ (Weinberg \&
Nikolaev~\cite{weinberg01}; Demers \& Battinelli~\cite{demers07}).
They are thus easily detectable at the distance of the Perseus arm
and beyond in a magnitude-limited sample like ours.

\subsection{Spectral classification in the visible\label{spec_vis}}

  The new spectroscopic classifications of 29 stars presented
for the first time in this paper\footnote{We inadvertently also
observed star \#044, already classified by Hanson~(\cite{hanson03}),
as noted in Table~\ref{tab_red_featureless}. For consistency, we use
here our own classification. Note in any case that the difference
between both (O9.7II in Hanson~(\cite{hanson03}), B0II according to
our own classification) is minimal.} are listed in Col.~7 of
Tables~\ref{tab_red_featureless} and \ref{tab_emission}.
Classifications were based on a comparison between our spectra and
the atlas of Walborn \& Kirkpatrick~(\cite{walborn90}), which
contains an abundant selection of early stars providing a very
finely-grained coverage of the two-dimensional classification of the
earliest spectral types, as well as an extensive discussion of the
classification criteria. In the range of spectral types covered by
our stars, spectral subtype classification is mainly based on the
relative strengths of the HeII features with respect to HeI, most
notably HeII~$\lambda$4541 vs. HeI~$\lambda$4471, for O-type stars
in which HeII lines are visible; and on the SiIV/SiIII and
SiII/SiIII ratios, as well as the HeI/H ratios, for the early B
stars. In turn, the luminosity class is largely derived from the
appearance of the lines produced by ionized metals (CIII, SiIV) in
the proximities of H$\gamma$ and, most notably, the intensity of the
CIII+OII feature at 4650~\AA. Because of the large extinction in
their direction, all our stars show prominent interstellar
absorption features, most notably the broad band centered at
4428~\AA \ and narrower lines at 4501~\AA, 4726~\AA, and 4762~\AA.
To estimate the uncertainty in the spectral classification, each
spectrum was separately classified by each of the authors using the
same reference atlas. A comparison of the individual results
obtained then led us to estimate an uncertainty in each dimension of
the spectral classification amounting to approximately one
luminosity class and less than one spectral subtype, respectively.
The spectral types listed in Tables~\ref{tab_red_featureless} and
\ref{tab_emission} are the averages of the classifications
attributed by each author.

\begin{figure*}
 \centering
 \includegraphics[width=16cm]{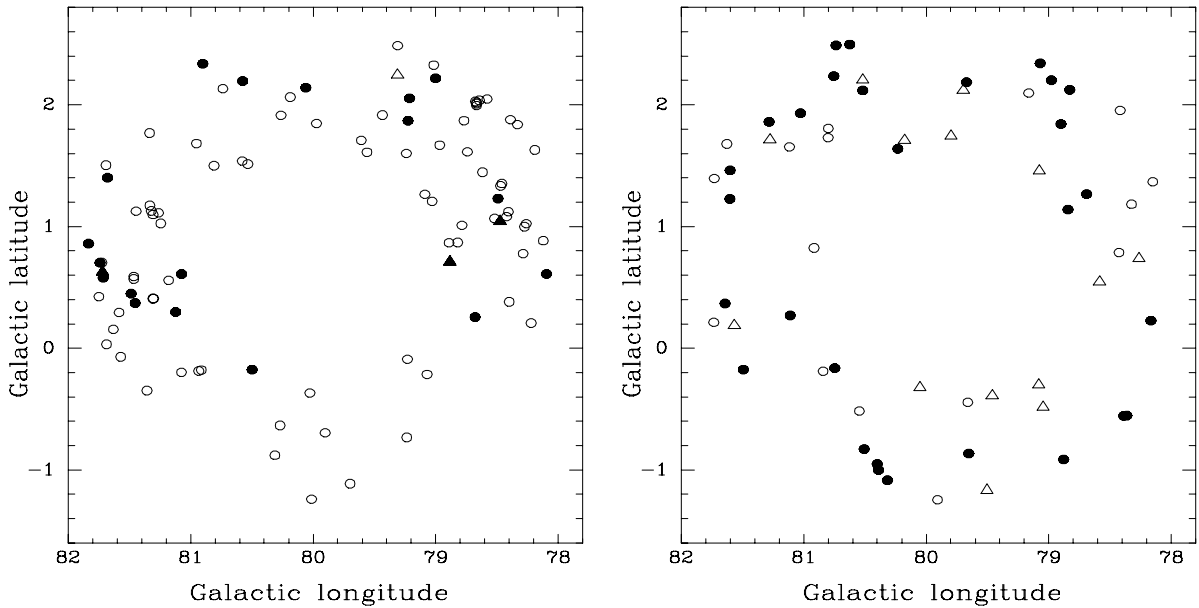}
 \caption[]{{\bf Left panel:} Location of all the candidate members
identified through their infrared spectra
(Tables~\ref{tab_red_featureless}, \ref{tab_emission},
\ref{tab_lightlyred_members}), including the protostellar candidates
listed in Table~\ref{tab_extremered} and star \#221. Symbols are as
follows: {\it open circles}, stars with featureless near-infrared
spectra; {\it filled circles}, stars with Brackett emission lines;
{\it filled triangles}, massive young stellar objects; and {\it open
triangle}, Wolf-Rayet star WR~142a. {\bf Right panel}: same for
three classes of non-members; {\it open circles}, red giant stars;
{\it filled circles}, carbon stars; and {\it open triangles},
long-period variables.}
  \label{fig_plot_pos}
\end{figure*}

  The spectral types obtained for the new members of the region are
encompassed within a very narrow range going from O7 to B2, with the
vast majority of the members having subtypes between O9 and B1. In
contrast, stars can be found in all the luminosity classes from I to
V. Our sample includes both objects with featureless spectra in the
near-infrared (Table~\ref{tab_red_featureless}) and objects observed
to display lines of the Brackett series in emission
(Table~\ref{tab_emission}). Since we restricted our visible
spectroscopy to targets with relatively bright magnitudes in the
blue, we only expect classical Be stars in the latter category to be
present in the sample observed in the visible. Indeed, none of the
stars that we observed in the visible displays the strong infrared
excess characteristic of Herbig Ae/Be stars (Hern\'andez et
al.~\cite{hernandez05}). On the other hand, of all the classical Be
star candidates only one, star \#128, also displays noticeable
emission in the visible in the form of a faint emission core at the
center of the Balmer lines.

\subsection{Spatial distribution\label{distribution}}

  Schematic plots with the spatial distributions of stars that we
consider as likely members and non-members of the Cygnus~OB2 halo
are presented in Fig.~\ref{fig_plot_pos}. Candidate members are
those listed in Tables~\ref{tab_red_featureless},
\ref{tab_emission}, \ref{tab_lightlyred_members},
\ref{tab_extremered}, as well as the young stellar object \#221.
Both populations have clearly different distributions: the
non-members are roughly uniformly spread out, as we may expect from
a population composed by old stars. Such uniform distribution
implies that there are no significant large-scale differences in
extinction along the line of sight at least out to the typical
distances of those stars, which as noted above are greater on the
average than those of early-type stars of the same apparent
magnitude.

\begin{figure}
 \centering
 \includegraphics[width=9cm]{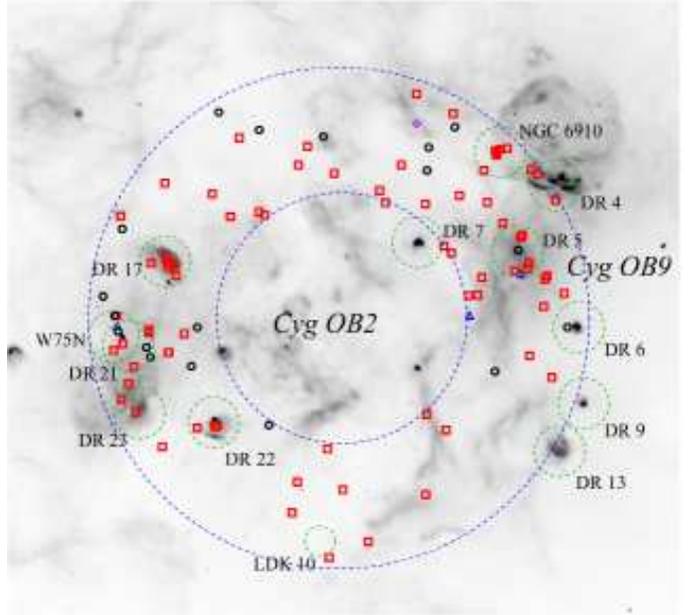}
 \caption[]{Location of all the 98 candidate early-type stars
overlaid on a radiocontinuum image of the region at 1420~MHz, from
the Canadian Galactic Plane Survey (Taylor et al.~\cite{taylor03}).
Stars are represented by the following symbols: {\it red squares},
stars with featureless near-infrared spectra; {\it black circles},
stars with Brackett emission lines; {\it blue triangles}, massive
young stellar objects; and {\it diamond}, Wolf-Rayet star WR~142a.
The large inner and outer blue dashed circles are respectively
$1^\circ$ and $2^\circ$ in radius, and they delimit the region in
the outskirts of Cygnus~OB2 covered by the present study. The
location of the main HII regions, as well as the W75N star forming
region and the NGC~6910 open cluster, are indicated. The HII regions
noted here with prefix 'DR' were identified by Downes \&
Rinehart~(\cite{downes66}). The prefix 'LDK' refers to the list of
embedded clusters of Le Duigou \& Kn\"odlseder~(\cite{leduigou02}).
The position of the compact HII region DR~7 is marked here, although
it is probably more distant and associated to the background Perseus
arm (Comer\'on \& Torra~\cite{comeron01})}.
  \label{fig_overlay}
\end{figure}

  The stars that we have considered as likely members, on the other
hand, do display structure in their spatial distribution. Although
OB stars are found all over the field, two areas of surface density
greater than the average appear toward the (galactic) northwest
(upper right in the left panel of Fig.~\ref{fig_plot_pos}) and
toward the east (left in that same figure).

  Figure~\ref{fig_overlay} identifies the main structures known in the
region, superimposed on a 1420~MHz radio continuum map from the
Canadian Galactic Plane Survey (Taylor et al.~\cite{taylor03})
tracing the emission by ionized gas. A comparison to
previously-known structures in that region shows that the
northeastern concentration corresponds to the eastern boundary of
the Cygnus~OB9 association (e.g., Garmany \&
Stencel~\cite{garmany92}), and in particular to the cluster NGC~6910
(the tight concentration of stars near $l = 78.7$, $b = +2.0$, which
are recovered by our candidate member selection procedure). On the
eastern side, the broad enhancement of stellar surface density
contains the chain of compact HII regions DR~17, DR~21 in the W75
complex, DR~23, and DR~22, all of which are embedded in the Cygnus~X
North complex. The reality of Cygnus~X as a single physical entity,
rather than as a chance alignment of unrelated structures lying at
different distances along a spiral arm seen nearly end-on, has been
widely debated in the literature (see, e.g., Odenwald \&
Schwartz~\cite{odenwald93}). Recent detailed observations of the
molecular gas in the region, mainly by Schneider et
al.~(\cite{schneider06}, \cite{schneider07}), strongly argue for a
real interconnection among these structures and present evidence for
their interaction with the massive stars of Cygnus~OB2. Based on
those results, we consider here that DR~17, W75/DR~21, DR~22, and
DR~23 are all structures belonging to Cygnus~X and lying at the same
distance as Cygnus~OB2, although we note that a distance of 2-3~kpc
to W75/DR21, significantly larger than the one adopted here, has
been often adopted in the literature (Kumar et al.~\cite{kumar07}).

  The concentration of early-type stars on the eastern side is
particularly obvious at the position of DR~17, composed by the
earliest-type members of the embedded clusters hosted by this HII
region (Le Duigou \& Kn\"odlseder~\cite{leduigou02}). However, it is
more widespread and extends beyond the boundaries of the compact HII
regions. A similar result has been reported recently by Kumar et
al.~(\cite{kumar07}), who find that many young stars revealed by
their infrared excesses in the Spitzer bands are located outside the
boundaries of the main molecular complexes in the region. We note,
in particular, the concentration of emission-line stars toward this
general region, with a preference toward the surroundings of DR~21,
although DR~21 itself is too obscured for its embedded stars to be
included in our magnitude-limited sample (Nadeau et
al.~\cite{nadeau91}). Almost all these stars have clear infrared
excess placing them among the Herbig Ae/Be stars as discussed in
Sect.~\ref{colcol}, with the only exception of star \#118, appearing
in the direction of W75N and whose near-infrared colors place it in
the classical Be star {\it locus}. Only another candidate classical
Be star is possibly associated with a compact HII region, namely
star \#122 near DR~6, although the spatial coincidence between them
is not precise.

  The southwestern (lower right) part of the area covered by this
study South of the galactic equator appears particularly devoid of
early-type stars. This is interesting since that region coincides
with Cygnus~X South (Schneider et al.~\cite{schneider06}), and
recent stellar surface density estimates based on
extinction-corrected starcounts (Bontemps et al., in prep;
preliminary results shown in Reipurth \&
Schneider~\cite{reipurth08}) indicate an enhanced stellar density in
that direction, which appears to be an extension of Cygnus~OB2. The
starcount enhancement detected by Bontemps et al. is contained
within $1^\circ$ from the center of Cygnus OB2 and is in rough
agreement with the distribution of early-type stars in the same
region found by Comer\'on et al.~(\cite{comeron02}; see Fig.~14 in
that paper), which preferentially extends toward the west. The same
distribution is well matched by that of A0V-A5V stars recently
identified by Drew et al.~(\cite{drew08}). As discussed by those
authors, the magnitudes of these stars suggest an average age
somewhat older than that of the Cygnus~OB2 cluster if they lie at
the distance of 1.45~kpc assumed here, in agreement with the results
of Hanson~(\cite{hanson03}), and it is most likely the lower mass
counterpart of the sample discussed by Comer\'on et
al.~(\cite{comeron02}). The fact that we detect no traces of it in
our color-selected sample beyond the $1^\circ$ circle suggests that
it either does not reach beyond that distance, or that its possible
extension does not contain massive stars picked up by our selection
criterion. On the other hand, $^{13}$CO maps of that region (Simon
et al., in prep; preliminary results shown in Reipurth \&
Schneider~\cite{reipurth08}) indicate the presence of large amounts
of molecular gas in the zone where we do not detect a corresponding
overdensity of early-type stars. It thus appears that any star
formation currently going on in that region of Cygnus~X is not
producing massive stars. It should be noted that this does not apply
to the entire Cygnus~X South region, which extends well beyond the
boundaries of the area considered here and does contain massive
embedded clusters and HII regions.

\subsection{Evolutionary status\label{evolutionary}}

  It is thus possible to identify three distinct components in the
area targeted by this study. The first one is the population
associated with the eastern edge of Cygnus~OB9, particularly the
cluster NGC~6910. The second component extends over the region
occupied by the DR~17, W75/DR~21, DR~22, and DR~23 complexes in
Cygnus~X North. The third component is a distributed population
whose members are stars that do not belong to either of those two
groups, which may be related to the extended population in the
central $1^\circ$-radius circle around Cygnus~OB2 reported by
Comer\'on et al.~(\cite{comeron02}) and Hanson~(\cite{hanson03}),
and probably also by Drew et al.~(\cite{drew08}).

  The new availability of detailed spectral classifications for many
stars in these three groups allows us to investigate in more detail
their nature and evolutionary status. To this purpose we have
adopted for each O-type star an effective temperature $T_{eff}$
based on the $T_{eff}$-spectral type calibration of Martins et
al.~(\cite{martins05}), using their observational effective
temperature scale. To our knowledge no similar work is available to
date covering the early B spectral interval, in particular the whole
range of luminosity classes covered by our spectra. We have thus
used the $T_{eff}$-spectral type compilation by
Tokunaga~(\cite{tokunaga00}) for luminosity classes I and V,
applying a scaling factor to the temperatures so as to force
agreement with Martins et al.'s~(\cite{martins05}) observational
temperature scale at spectral types O9I and O9.5V respectively. The
calibrations of Tokunaga~(\cite{tokunaga00}) and Martins et
al.~(\cite{martins05}) closely match each other in the overlapping
late-O range, implying that our arbitrary scaling factors deviate
from unity by less than 4\%. Because the compilation of
Tokunaga~(\cite{tokunaga00}) does not include luminosity class III
early-type stars, we have estimated their temperatures by means of a
weighted interpolation between the temperatures adopted for classes
I and V. The weight factor is derived once again by forcing
agreement between the temperatures derived in this way for spectral
type O9III and the corresponding value given by Martins et
al.~(\cite{martins05}). Approximate temperatures of B stars of
luminosity classes II and IV have then been computed by averaging
the temperatures adopted for stars of luminosity class I and III,
and III and V, respectively. We estimate that the systematic errors
introduced in this way are below 2000~K for all the spectral types
represented in our sample. A summary of $T_{eff}$-spectral type
calibrations for B-type stars presented by Fitzpatrick \&
Massa~(\cite{fitzpatrick05}) shows that existing systematic
uncertainties in this domain exceed this value. We thus consider
that the {\it ad-hoc} procedure that we have described to assign
effective temperatures to our stars does not introduce large
additional uncertainties.

  The large scatter in absolute magnitudes at a given spectral type
among O and early-B stars (e.g., Jaschek \&
G\'omez~\cite{jaschek98}) cautions against the use of the spectral
classification alone to derive the positions of our stars in the
temperature-absolute magnitude diagram, and compare them to the
predictions of evolutionary tracks to assess their evolutionary
status. Instead, we have preferred to adhere to the underlying
hypothesis already noted in Sect.~\ref{distribution} that all our
stars have essentially the same distance modulus of 10.80~mag, and
derive individual absolute magnitudes as

$$M_V = K - 10.80 - 0.659[(J-K)-(J-K)_0] + (V-K)_0$$

\noindent where $(J-K)_0$ and $(V-K)_0$ are the intrinsic,
unreddened color indices and the factor 0.659 is derived from the
extinction law of Rieke \& Lebofsky~(\cite{rieke85}). We have used
the intrinsic colors derived by Martins \& Plez~(\cite{martins06})
for O stars and those compiled by Tokunaga~(\cite{tokunaga00}) for B
stars. We estimate that interpolations needed to cover all the
spectral types and luminosity classes in our sample introduce a
systematic error in $M_V$ not exceeding 0.1~mag. The scatter in
intrinsic colors among actual stars is likely to be a more important
source of errors, as illustrated by the results of Robert et
al.~(\cite{robert03}).

\begin{figure*}
 \centering
 \includegraphics[width=18cm]{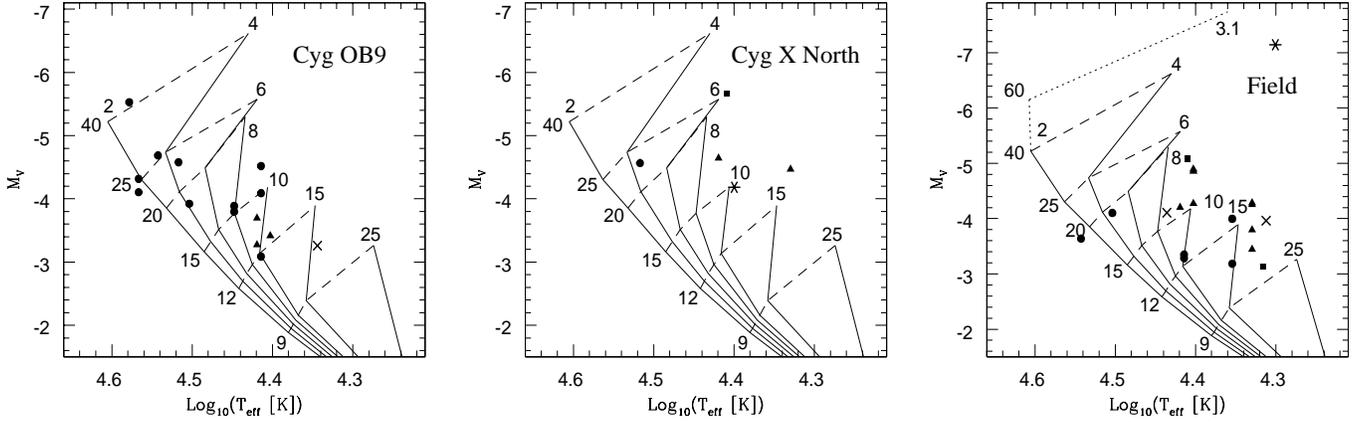}
 \caption[]{Temperature-absolute magnitude diagram with the location
of the stars for which spectral classification is available in each
of the three regions discussed in Sect.~\ref{evolutionary}. Circles
indicate luminosity class V, crosses class IV, triangles class III,
squares class II, and asterisks class I. The evolutionary tracks and
isochrones from Lejeune and Schaerer~(\cite{lejeune01}) are plotted
for comparison, highlighting the older age of the field population.
We believe that most of the stars in the Cygnus X North area
actually belong to the part of the field population lying along that
line of sight.}
  \label{fig_hr}
\end{figure*}

  The T$_{eff}$-$M_V$ diagrams are shown in Fig.~\ref{fig_hr} for the
eastern edge of Cygnus~OB9, the region in Cygnus~X North around
DR~21, and the field population. In these figures, we also plot the
isochrones calculated by Lejeune \& Schaerer~(\cite{lejeune01}) in
their Case $e$ with solar metallicity, high mass loss and no
rotation, for initial stellar masses (60), 40, 25, 20, 15, 12 and 9
M$_{\odot}$ (solid lines) and stellar ages 2, (3.1), 4, 6, 8, 10, 15
and 25 Myr (dotted lines). Individual $T_{eff}$ and $M_V$ derived
for each star in the different regions are listed in
Table~\ref{tab_teff_mv}.

\onltab{10}{
\begin{table}
\caption[]{Adopted temperatures and absolute
magnitudes\label{tab_teff_mv}}
\begin{tabular}{clcc}
\hline
Star & Sp. type & $M_V$ & $\log T_{eff} {\rm (K)}$ \\
\hline \noalign{\medskip {\it Cygnus OB9 area}
\medskip}
27  & O6.5V   & -5.53 & 4.578 \\
62  & O8V     & -4.69 & 4.542 \\
68  & B1V     & -4.52 & 4.414 \\
82  & O9V     & -4.58 & 4.517 \\
101 & O7V     & -4.32 & 4.567 \\
108 & B1V     & -4.09 & 4.414 \\
137 & B0.5V   & -3.80 & 4.448 \\
138 & B0.5V   & -3.89 & 4.448 \\
142 & B1.5IV  & -3.26 & 4.344 \\
159 & O9.5V   & -3.92 & 4.504 \\
164 & O7V     & -4.10 & 4.567 \\
181 & B0III   & -3.69 & 4.419 \\
193 & B0.2III & -3.42 & 4.403 \\
206 & B0III   & -3.28 & 4.419 \\
232 & B1V     & -3.09 & 4.414 \\
\noalign{\medskip {\it Cygnus X North (DR 21) area}
\medskip}
36  & B0II    & -5.67 & 4.409 \\
65  & B1III   & -4.47 & 4.329 \\
97  & B0III   & -4.65 & 4.419 \\
100 & O9V     & -4.57 & 4.517 \\
120 & B0Ib    & -4.19 & 4.399 \\
\noalign{\medskip {\it Field area}
\medskip}
10  & B1Ib      & -7.14 & 4.301 \\
44  & B0II      & -5.08 & 4.409 \\
45  & B0.2III   & -4.85 & 4.403 \\
50  & B3II      & -4.03 & 4.191 \\
60  & B0.2III   & -4.89 & 4.403 \\
64  & B0.2III   & -4.27 & 4.403 \\
71  & B2IV      & -3.96 & 4.312 \\
94  & B1III     & -4.28 & 4.329 \\
110 & B1III     & -4.25 & 4.329 \\
112 & B2V       & -4.00 & 4.354 \\
122 & B0.2IV    & -4.11 & 4.436 \\
128 & B0IIIe    & -4.21 & 4.419 \\
133 & B1III     & -3.80 & 4.329 \\
150 & O8V       & -3.64 & 4.543 \\
152 & O9.5V     & -4.10 & 4.504 \\
175 & B1III     & -3.45 & 4.329 \\
196 & B1V       & -3.28 & 4.414 \\
218 & B2V       & -3.18 & 4.354 \\
229 & B1V       & -3.35 & 4.414 \\
244 & B1II      & -3.13 & 4.315 \\
\hline\smallskip
\end{tabular}\\
\end{table}
}

  There are significant differences between the distributions of
stars in the temperature-luminosity diagrams among the three
components considered. The Cygnus~OB9 region contains abundant
O-type main sequence stars, particularly in the cluster NGC~6910
whose members display other signatures of youth (Delgado \&
Alfaro~\cite{delgado00}), and only four stars outside the main
sequence (the B giants \#181, 193, and 206 and the B1.5IV star
\#142) are found within this region in our sample. The class V
O-type stars appear mostly above the main sequence, which may
indicate that they lie at a shorter distance than that adopted by
us. Garmany et al.~(\cite{garmany92}) find a very uncertain distance
modulus $DM = 10.0$ to NGC~6910, in contrast with the better
determined value for the rest of the association, $DM = 11.0$.
However, this latter value is in closer agreement with the distance
found by Delgado \& Alfaro~(\cite{delgado00}) for NGC 6910 itself,
$DM = 11.2 \pm 0.2$, which solves the discrepancy noted by Garmany
\& Stencel~(\cite{garmany92}) and is only marginally larger than
that adopted by us. A value significantly shorter than $DM = 10.8$
thus does not seem supported by other observations. The position of
the class V O-type stars above the lowest isochrone may be at least
partly due to binarity, as well as to rotation, whose effects can
brighten O stars by as much as $\Delta M_V = 0.6$ (Meynet \&
Maeder~\cite{meynet00}) for an initial velocity of 200~km~s$^{-1}$.

  The situation is less clear in Cygnus~X North, as
the higher extinction in the area prevented us from obtaining
visible spectra of many stars. The presence of several Brackett
emission-line stars with the infrared characteristics of Herbig
Ae/Be stars suggests recent star formation over a widespread area.
The five stars with spectral classification in the zone are all
lightly reddened and may be on the near side of the complex. The
luminosity classes (Ib-III) of four of them indicates that they are
not related to the ongoing star formation in the area, with the
possible exception of the O9V star \#100 that is very close to
DR~21. It can be seen in Fig.~\ref{fig_hr} that a physical
association of this star with DR~21 would support the distance
adopted here, rather than the 2-3~kpc often used in the literature
(Davis et al.~\cite{davis07}), as a greater distance would move it
even further away from the main sequence.

  In contrast, most of the population outside those two areas, which
we consider as the field component, is lightly reddened, and visible
spectroscopy is available for 20 of its members, of which 14 are
found to have luminosity classes placing them above the main
sequence. Of the remaining six stars, four are classified as B1V or
B2V and thus main sequence lifetimes longer than 10~Myr; and only
three stars, \#150 (O8V; Morgan et al.~\cite{morgan55}), \#152
(O9.5V; this work) and WR~142a (WC8, for which Pasquali et
al.~(\cite{pasquali02}) estimate an age of at least 3~Myr) indicate
more recent massive star formation. The first two are also the ones
closest to the main sequence plotted on Fig.~\ref{fig_hr}, in
agreement with the common distance adopted for all our stars.
Besides WR~142a, 7 out of 9 emission line stars in the region appear
to be classical Be stars, with the only exceptions of stars \#178
and \#180.  A comparison of the position of the stars belonging to
the field component with the evolutionary tracks in the
temperature-luminosity diagram indicates ages older than 6~Myr for
this population.

  It may be noted that several stars classified as B1V or B2V appear
up to 2~mag above the main sequence in the temperature-magnitude
diagrams for both Cygnus~OB9 and the field population. This is
considerably more than what one may expect from the effects of
rotation, binarity, or the spread in intrinsic magnitudes within
their spectral types. In fact such stars are not expected to be
present in our sample, as normal main sequence stars later than B1V
at the distance of Cygnus~OB2 should have $K_S$ magnitudes fainter
than our target selection limit (see Sect.~\ref{target_selection}).
In principle, these may be unrelated foreground stars at much closer
distances. However, an inspection of their spectra in
Fig.~\ref{fig_specvis} shows interstellar features of a depth
similar to those of the other stars in the region, indicating a
similar level of extinction that is confirmed by their infrared
photometry. A misclassification of the luminosity class thus appears
as a more likely cause, which may be confirmed with better quality
spectra. However, we have decided not to modify the spectral types
given in this paper so as not to bias our classification with {\it a
posteriori} knowledge.

\section{Discussion\label{discussion}}

  Our observations sample an interesting region where the outskirts
of the Cygnus~OB2 association merge with neighbor structures,
allowing us to explore the possible existence of links among them
and the history of star formation in the whole region. Such
questions are rendered all the more relevant given the increasing
evidence for their common distance (Schneider et
al.~\cite{schneider07}) and the possibility that the distinct
structures recognized thus far are parts of a single association,
particularly Cygnus~1, 8, and 9 (Mel'nik \&
Efremov~\cite{melnik95}).

  One of our main results is the identification
of an extended distribution of OB stars around Cygnus~OB2 and
probably beyond, selected through their near-infrared colors and
confirmed on the basis of their near-infrared spectrum. Visible
spectroscopy confirms the very high reliability of infrared
spectroscopy at modest resolution and signal-to-noise ratio in
identifying early-type stars, as already found by
Hanson~(\cite{hanson03}) when investigating the sample of candidates
identified by Comer\'on et al.~(\cite{comeron02}). Our results also
show a remarkable coincidence with those of Hanson~(\cite{hanson03})
in that the newly-identified early-type stars do not have, as a
rule, spectral types as early as those at the center of Cygnus~OB2.
Instead, they are dominated by massive stars that have already
evolved off the main sequence (i.e., giants, supergiants, and a
Wolf-Rayet) or are thought to be at the end of their main sequence
lifetime (the classical Be stars). Such evolved stars are dominant
among the field component described in Sect.~\ref{evolutionary}.
Their presence also among the populations that we identified as
belonging to the complex that includes DR~17, DR~21, DR~22, and
DR~23, on one side, and Cygnus~OB9, on the other, is not surprising
since our allocation of stars to those groups is purely based on
position. A certain level of contamination of those two components
by stars actually belonging to the field population is thus to be
expected.

  What is the relationship between the field component and Cygnus~OB2,
if there is one? Hanson~(\cite{hanson03}) already pointed out that
the spatial distribution of the evolved component differed markedly
from that of the most massive stars of Cygnus~OB2 in being much more
evenly distributed over the studied area. In fact, based on both the
wider spatial distribution and the gap in age Hanson suggested that
the evolved component is unrelated to Cygnus~OB2. Our new
observations allow us to reevaluate this conclusion by extending the
analysis of the spatial distribution to a larger area.

  Unfortunately, the somewhat different selection of candidate
early-B stars carried out in Comer\'on et al.~(\cite{comeron02}) and
the present paper prevents a direct comparison between the stellar
densities inside and outside the $1^\circ$ radius. It is however
possible to perform the color selection of candidates over a much
larger area that includes the central 1$^\circ$ radius around
Cygnus~OB2, the region covered by this study, and the surrounding
regions, using the results of Sect.~\ref{class_ir} as a way to
remove statistically the contamination by cool background stars and
by foreground stars.

\begin{figure}
 \centering
 \includegraphics[width=9cm]{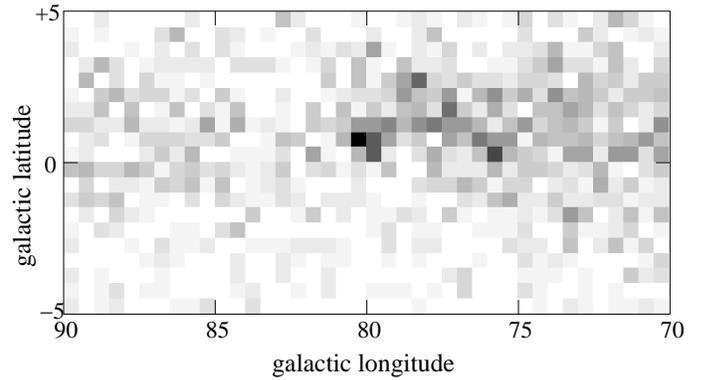}
 \caption[]{Surface density of OB stars selected according to the criteria
$K < 8.8$, $-0.15 < (J-H)-1.70 (H-K_S) < 0.20$. The shade of grey
scales with the number of stars per square degree, ranging from zero
(white) to 100 or more (black). The Cygnus~OB2 central cluster is
the black square near the center, with a density of 176 stars per
square degree. A pedestal surface density of 12.5 stars per square
degree has been subtracted from each cell to account for the
estimated contamination by cool and foreground stars that are also
selected by the color-based criteria.}
  \label{fig_surfdens}
\end{figure}

  We have carried out this exercise over the area limited by
$70^\circ < l < 90^\circ$, $-5^\circ < b < +5^\circ$, which has
Cygnus~OB2 near its center and includes the entire Cygnus~OB9 region
as well as parts of Cygnus~OB1 to the West and Cygnus~OB7 to the
East. Stars were selected according to the criteria $K_S < 8.8$,
$-0.15 < (J-H)-1.70 (H-K_S) < 0.20$. The results of the
spectroscopic follow-up discussed in Sect.~\ref{class_ir} indicate
that these criteria select approximately 12.5 cool (red giants,
long-period variables, and carbon stars) and foreground (mostly late
B to F-type) stars per square degree, in addition to OB stars. We
assume that these contaminants are uniformly distributed across the
field, and subtract this number density from the total number
density of selected objects. The result is displayed in
Fig.~\ref{fig_surfdens}, which is similar to the surface density
maps presented by Kn\"odlseder~(\cite{knoedlseder00}) and Bontemps
et al. (in prep.; presented by Reipurth \&
Schneider~\cite{reipurth08}). The main difference is that
Fig.~\ref{fig_surfdens} is restricted to early-type stars; the
surface densities are thus much lower and the resolution much
coarser than in their contour maps. As one may expect, Cygnus~OB2 is
the most prominent feature appearing as a very tight concentration
near the center of the figure, with a peak surface density reaching
176 OB stars per square degree in the central 0.25 square degrees.
The stellar density remains high toward the western half occupied by
Cygnus~OB9 next to Cygnus~OB2, and then Cygnus~OB1. The easternmost
quarter of the figure is occupied by the more nearby association
Cygnus~OB7 (Garmany \& Stencel~\cite{garmany92}). Despite the
overall similarity, there are significant differences between our
map and the general starcount-based map of Bontemps et al.:
Cygnus~OB9 appears much more prominently in our map, whereas the
DR~17/DR~21/DR~22/DR23 region, while easily recognizable as a
stellar density enhancement in our Fig.~\ref{fig_overlay}, stands
out much more clearly in Bontemps et al.'s map.

  Interestingly, Fig.~\ref{fig_surfdens} shows that a non-zero
surface density of early-type star candidates pervades the region
even outside the boundaries of the already-known OB associations. By
extrapolating the results obtained in the area between $1^\circ$ and
$2^\circ$ from the center of Cygnus~OB2, we tentatively identify
this as the field population mainly composed of evolved stars,
noting that this population does not appear to be associated with
Cygnus~OB2 nor with any other specific feature in the region. The
massive precursors required by this widespread field population
suggests that massive star formation has been taking place in the
Cygnus region for a time exceeding 10~Myr, and long before the
formation of the currently observed associations like Cygnus~OB2,
whose age is less than 4~Myr (Massey et al.~\cite{massey95}). The
fact that such a population appears as an extended component with no
apparent concentration is not entirely surprising, given the high
density of OB associations in the region, if we assume that this
density was similarly high in the past. The typical center-to-center
distance among the four associations currently defined in the
galactic longitude interval $74^\circ < l < 82^\circ$ (Cygnus~OB1,
OB2, OB8, and OB9) is $\sim 2^\circ 5$. At the distance of 1.45~kpc
adopted throughout this paper, an internal velocity dispersion of
3.5~km~s$^{-1}$ would suffice to make them overlap after 10~Myr if
they are gravitationally unbound, even if they occupied a small
initial volume. This velocity dispersion is comparable to the
internal velocity dispersion derived for members of Cygnus~OB2 by
Kiminki et al.~(\cite{kiminki07}), $2.44 \pm 0.07$~km~s$^{-1}$.
Moreover it can be regarded as an upper limit, since even very young
OB associations like Cygnus~OB9 occupy a sizeable volume comparable
to the center-to-center angular distance quoted above, and the field
population contains members older than 10~Myr leaving more time for
dispersal of the original associations. This population may be the
equivalent to an extended component also observed in other galaxies
(Pasquali \& Castangia~\cite{pasquali08}), and it can thus be
accounted for by assuming a previous generation of OB associations,
now dispersed, with characteristic similar to those of the current
associations. It appears unlikely that the field population has its
origin in bound clusters having become unbound in their early
evolutionary stages (Lada \& Lada~\cite{lada03}): recent
observational evidence discussed by Gieles \&
Bastian~(\cite{gieles08}) indicates dispersal timescales at least
one order of magnitude longer than the age of the field population
in a wide range of environments, including our galactic
neighborhood. Our results rather suggest that those massive OB stars
originate in either isolation or unbound aggregates.

  The sustained massive star forming activity in Cygnus must have been
essential in arranging the large-scale distribution of the
interstellar gas and in driving the current generation of star
formation. It may also be at the origin of the large Cygnus
Supperbubble observed in X rays (Cash et al.~\cite{cash80},
Bochkarev \& Sitnik~\cite{bochkarev85}, Uyaniker et
al.~\cite{uyaniker01}) and possibly related phenomena such as the
apparent large-scale expansion pattern noted by Comer\'on et
al.~(\cite{comeron98}) in the proper motions of stars in the area.

\section{Conclusions\label{conclusions}}

  We have extended previous work carried out mainly
by Comer\'on et al.~(\cite{comeron02}) and Hanson~(\cite{hanson03})
in characterizing the population of early-type stars in the vicinity
of Cygnus~OB2. We have done this by identifying a magnitude-limited
sample of candidate early-type stars through near-infrared imaging
in the area between $1^\circ$ and $2^\circ$ from the center of
Cygnus~OB2, confirming them through near-infrared spectroscopy, and
providing an accurate spectral classification of many of them
through spectroscopy in the visible. Our main results can be
summarized as follows:

\begin{itemize}

\item The studied area is rich in early-type stars, including
evolved massive emission-line stars, likely Herbig Ae/Be stars, and
some deeply-embedded massive young stellar objects. As in previous
studies, a color-based selection criterion is found to be efficient
in selecting candidate early-type stars, although with a
considerable degree of contamination by other types of objects,
mainly cool stars and foreground late-B, A, and F stars.
Near-infrared spectroscopy at modest resolution and signal-to-noise
ratio suffices to identify the actual emission-line stars.

\item Our early-type selection criteria recovers many stars
belonging to the Cygnus~OB9 association and its cluster NGC~6910,
which are partly included in the studied area. We also find a
significant number of reddened early-type stars and Herbig Ae/Be
star candidates in the area occupied by the Cygnus~X North complex,
which includes the compact HII regions DR~17, DR~21, DR~22, and
DR~23. The massive star distribution is, however, not confined to
the HII regions.

\item Many early-type stars are found not to be associated with any
known structure. Visible spectroscopy is available for most of this
field population, and shows that it is strongly dominated by giant
and supergiant early-B stars. The characteristics of this population
appear to be the same as discussed by Hanson~(\cite{hanson03}) for
the additional early-type stars identified by Comer\'on et
al.~(\cite{comeron02}) in the vicinity of Cygnus~OB2. It most
probably represents a population significantly older than the
massive stellar cluster at the center of Cygnus~OB2.

\item We have extended the color-based selection criterion to
estimate the surface density of early-type candidates in an area of
200~square degrees centered on Cygnus~OB2, statistically correcting
for the unrelated background and foreground populations. We find
that the field population is not particularly related to Cygnus~OB2
nor to any other structure of the region, being all-pervasive. The
need for massive precursors to produce this extended field
population suggests that massive star formation in the whole Cygnus
region has proceeded for a long time, well before the currently
identified OB associations were formed. Such interpretation rests on
the assumption underlying this work that all the early-type stars
are located at similar distances, and it would be challenged if the
field population were either foreground or background to the OB
associations, or if it spread over a wide range of distances.
Although this possibility needs to be clarified by future studies,
we do not find evidence in our temperature-absolute magnitude
diagrams for a systematic difference between the distance to the
members of this population and the adopted distance of Cygnus~OB2.

\end{itemize}

  We believe that the present study conclusively dismisses the case
for a large extent of Cygnus~OB2 much beyond the boundaries of its
central concentration, which was already suggested by Garmany \&
Stencel~(\cite{garmany92}) and supported by Comer\'on et
al.~(\cite{comeron02}), but then questioned by
Hanson~(\cite{hanson03}). Previous studies such as those by Massey
\& Thompson~(\cite{massey91}) and
Kn\"odlseder~(\cite{knoedlseder00}) clearly show that most of the
massive stellar content of the association still awaits
identification and classification (see also Kiminki et
al.~\cite{kiminki07}, for recent work on the confirmation of new OB
members in the inner regions of Cygnus~OB2). Our results indicate
that future studies of Cygnus~OB2 aiming to characterize its upper
main sequence should concentrate on heavily-reddened members near
its core, rather than on identifying new members at large distances
from it.

  We note that over half of the early-type stars for which we have
obtained confirming near-infrared spectroscopy do not yet have a
detailed spectral classification available. This is particularly
important for the new candidates identified in Cygnus~X North, whose
proximity to compact HII regions, some of which contain embedded
clusters, leads us to suspect a young age for most of them. If
confirmed, the extent and composition of this group may warrant its
status as an OB association in its own right, although we feel that
a better characterization is still needed before taking that step.
On the other hand, the as yet unclassified stars located elsewhere
offer a potential for the discovery of new interesting members
deserving detailed study, such as WR~142a or BD$+43^\circ 3654$. The
fact that the vast majority of stars that form the basis for the
present study are recognized here as early-type members of the
region for the first time is a reminder of the large amount of
observational work still needed to characterize the stellar
population in the Cygnus region. The importance of this task is
stressed by the rapid progress being made at different wavelengths
in the characterization of its interstellar content.

\begin{acknowledgements}
  Once again it is a pleasure to thank the staff of the Calar Alto
observatory, and especially Santos Pedraz and Ana Guijarro, for
their unfailingly competent and friendly support while observing on
Calar Alto. We are very thankful to Bo Reipurth and Nicola Schneider
for making available to us their chapter on the Cygnus region for
the Handbook of Star Forming Regions prior to publication. We also
thank the anonymous referee for constructive comments that helped
improve both the style and content of this paper. FC warmly
acknowledges the hospitality of the Vatican Observatory during the
preparation of this paper. AP acknowledges support from the OPTICON
Network. JT and FF acknowledge support by the Spanish Ministry of
Science and Technology under contract AYA2006-15623-C02-02. This
research has made use of the SIMBAD database operated at CDS,
Strasbourg, France. It also makes use of data products from the Two
Micron All Sky Survey, which is a joint project of the University of
Massachusetts and the Infrared Processing and Analysis
Center/California Institute of Technology, funded by the National
Aeronautics and Space Administration and the National Science
Foundation; and of data from the Canadian Galactic Plane Survey, a
Canadian project with international partners, supported by the
Natural Sciences and Engineering Research Council.

\end{acknowledgements}

\onllongtab{1}{
\begin{longtable}{cccccccll}
\caption{Stars with reddened, featureless spectra\label{tab_red_featureless}}\\
\hline\hline
Star & $\alpha (2000)$ & $\delta(2000)$ & $J$ & $H$ & $K_S$ & Vis. classification & Other IDs & Notes \\
\hline
\endfirsthead
\caption{continued.}\\
\hline\hline
Star & $\alpha (2000)$ & $\delta(2000)$ & $J$ & $H$ & $K_S$ & Vis. classification & Other IDs & Notes \\
\hline
\endhead
\hline
\endfoot
  10 & 20  31  42.15 & 42  25  53.2 &   6.212 & 5.467 &  5.039 & B1Ib & & \\
  33 & 20  25  27.27 & 40  24  00.1 &   7.200 & 6.852 &  6.652 & & \object{BD+39 4179} & \\
  36 & 20  38  20.40 & 41  56  56.4 &   7.839 & 7.086 &  6.686 & B0II & & \\
  37 & 20  39  15.58 & 42  17  10.9 &   8.870 &  7.554 &  6.830 & & & 1 \\
  39 & 20  35  08.66 & 42  25  57.0 &   9.359 & 7.801 &  6.966 & & & 2 \\
  42 & 20  35  42.96 & 42  29  41.9 &   8.222 & 7.426 &  6.993 & & & 3 \\
  44 & 20  31  08.39 & 42  02  42.3 &   7.815 & 7.298 &  7.029 & B0II & \object{A41}& 4\\
  45 & 20  32  02.20 & 42  12  26.1 &   7.534 & 7.264 &  7.049 & B0.2III & \object{BD+41~3794} & \\
  60 & 20  27  52.95 & 41  44  06.8 &   8.152 & 7.580 &  7.271 & B0.2III & & \\
  62 & 20  23  22.84 & 40  09  22.5 &   7.581 & 7.421 &  7.286 & & \object{HD 229202} & 5 \\
  65 & 20  35  10.62 & 42  20  22.6 &   7.829 & 7.510 &  7.326 & B1III & \object {LS III +42 17} & 6 \\
  70 & 20  39  24.00 & 40  29  39.3 &   8.734 & 7.862 &  7.373 & & & \\
  72 & 20  26  10.44 & 39  51  51.9 &   8.745 & 7.847 &  7.388 & & & \\
  74 & 20  32  33.51 & 42  47  25.5 &   8.963 & 7.977 &  7.443 & & & \\
  75 & 20  41  05.74 & 39  55  19.5 &   8.526 & 7.806 &  7.448 & & & \\
  82 & 20  25  06.52 & 40  35  49.6 &   7.931 & 7.650 &  7.490 & O9V & \object{BD+40 4159} & \\
  93 & 20  40  29.78 & 42  03  13.7 &   9.243 & 8.156 &  7.619 & & & 7 \\
  94 & 20  25  33.19 & 40  48  44.5 &   8.337 & 7.900 &  7.639 & B1III & & \\
  96 & 20  26  46.79 & 40  05  57.0 &   8.955 & 8.073 &  7.655 & & & 8 \\
  97 & 20  38  21.72 & 41  57  06.8 &   8.755 & 8.037 &  7.680 & B0III & & \\
 100 & 20  39  45.06 & 42  06  07.8 &   8.470 & 7.984 &  7.712 & O9V & & \\
 101 & 20  27  17.57 & 39  44  32.6 &   8.100 & 7.822 &  7.727 & & \object{LS II +39 53} & 9\\
 108 & 20  23  14.54 & 40  45  19.3 &   8.157 & 7.932 &  7.783 & B1V & \object{BD+40~4149} & 10\\
 110 & 20  23  02.91 & 41  33  46.7 &   8.758 & 8.140 &  7.825 & B1III & & \\
 129 & 20  29  29.76 & 39  21  22.4 &   9.538 & 8.523 &  8.003 & & & \\
 130 & 20  40  33.68 & 40  22  45.1 &   9.775 & 8.606 &  8.028 & & & \\
 131 & 20  28  45.54 & 40  17  04.0 &   9.239 & 8.456 &  8.043 & & & \\
 132 & 20  37  17.74 & 41  56  31.4 &   9.092 & 8.379 &  8.053 & & & \\
 133 & 20  27  49.25 & 40  17  00.4 &   8.705 & 8.310 &  8.079 & B1III & & \\
 134 & 20  39  44.16 & 42  18  42.6 &   9.534 & 8.580 &  8.094 & & & 11 \\
 135 & 20  33  52.19 & 39  47  26.5 &   9.271 & 8.482 &  8.105 & & & \\
 137 & 20  23  07.58 & 40  46  09.0 &   8.480 & 8.244 &  8.126 & & & 10, 12 \\
 138 & 20  24  05.15 & 40  46  03.6 &   8.621 & 8.310 &  8.127 & B0.5V & \object{LS~III~+40~15} & \\
 140 & 20  39  34.10 & 41  16  56.5 &  10.080 & 8.800 &  8.140 & & & 13 \\
 143 & 20  35  22.96 & 42  22  02.4 &   9.417 & 8.613 &  8.155 & & & 14 \\
 146 & 20  23  07.30 & 40  46  55.2 &   8.504 & 8.254 &  8.206 & & & 10\\
 152 & 20  27  37.87 & 41  15  46.8 &   9.172 & 8.581 &  8.264 & O9.5V & & \\
 156 & 20  41  15.71 & 41  52  06.9 &  10.218 & 8.988 &  8.289 & & & 15\\
 159 & 20  26  24.88 & 40  01  41.4 &   9.044 & 8.569 &  8.322 & O9.5V & & 16 \\
 164 & 20  26  19.75 & 39  51  42.5 &   9.344 & 8.723 &  8.349 & O7V & & \\
 169 & 20  39  40.24 & 41  17  34.0 &  10.329 & 9.098 &  8.389 & & & 17 \\
 175 & 20  29  02.47 & 42  31  16.0 &   9.056 & 8.626 &  8.428 & B1III & & \\
 181 & 20  26  11.89 & 40  02  22.4 &   9.282 & 8.746 &  8.464 & B0III & & 18 \\
 183 & 20  29  18.68 & 39  36  15.8 &   9.793 & 8.910 &  8.471 & & & 19 \\
 185 & 20  22  50.95 & 41  13  50.7 &   9.785 & 8.973 &  8.519 & & & \\
 186 & 20  34  52.07 & 42  55  08.1 &   9.615 & 8.901 &  8.524 & & & \\
 189 & 20  33  50.76 & 39  59  38.7 &   9.675 & 8.925 &  8.541 & & & \\
 190 & 20  41  12.17 & 42  01  23.7 &  10.726 & 9.332 &  8.548 & & & 20 \\
 193 & 20  22  54.49 & 40  23  31.7 &   9.140 & 8.753 &  8.553 & B0.2III & & \\
 196 & 20  27  39.82 & 40  40  38.6 &   8.938 & 8.673 &  8.579 & B1V & & \\
 202 & 20  28  31.87 & 40  13  42.7 &  10.601 & 9.299 &  8.613 & & & \\
 204 & 20  35  29.68 & 42  16  24.3 &  10.669 & 9.308 &  8.618 & & & 21 \\
 205 & 20  35  18.14 & 42  23  48.4 &  10.052 & 9.075 &  8.620 & & & 22 \\
 206 & 20  25  28.89 & 40  12  54.2 &   9.058 & 8.750 &  8.627 & B0III & \object{LS~III +40~32} & 23 \\
 209 & 20  39  34.16 & 39  45  02.3 &  10.015 & 9.116 &  8.634 & & & \\
 219 & 20  22  54.30 & 40  27  41.2 &   9.085 & 8.806 &  8.700 & & \object{HD~229167} & \\
 220 & 20  36  32.51 & 39  36  47.2 &  10.054 & 9.189 &  8.713 & & & 24 \\
 222 & 20  37  30.82 & 40  27  52.0 &  10.580 & 9.351 &  8.723 & & & \\
 227 & 20  38  05.37 & 42  11  14.2 &   9.964 & 9.162 &  8.740 & & & \\
 229 & 20  27  20.99 & 41  21  26.2 &   9.460 & 8.981 &  8.746 & B1V & & \\
 230 & 20  41  43.65 & 41  31  45.2 &  10.090 & 9.224 &  8.767 & & & \\
 232 & 20  22  58.94 & 40  45  39.4 &   9.131 & 8.905 &  8.774 & B1V & & 25 \\
 242 & 20  38  26.96 & 42  27  44.8 &   9.876 & 9.145 &  8.823 & & & 26 \\
 244 & 20  31  06.24 & 41  59  32.4 &   9.615 & 9.128 &  8.824 & B1II & & \\
 246 & 20  26  40.87 & 40  59  46.9 &  11.198 & 9.652 &  8.829 & & & \\
 250 & 20  40  10.22 & 41  23  53.5 &  10.901 & 9.565 &  8.852 & & & 27 \\
\end{longtable}
{\it Notes:} \smallskip\\
1: In DR~21, possibly associated to the radiosource \object{WSRTGP
2037+4206} (Taylor et al.~\cite{taylor96}).\\
2: In DR~17.
3: In DR~17. Brightest star in cluster 14 of Le Duigou \& Kn\"odlseder~(\cite{leduigou02})\\
4: In Comer\'on et al.~(\cite{comeron02}); classified as O9.7II by
Hanson~(\cite{hanson03}).\\
5: Classified as O8V: by Morgan et al.~(\cite{morgan55}); X-ray emission (\cite{chlebowski89})\\
6: Possibly related to the HII region \object{G081.3+01.1} in
\object{DR17}.\\
7: In DR~23. Brightest star in cluster 15 of Le Duigou \&
Kn\"odlseder~(\cite{leduigou02}).\\
8: Possibly associated to DR~5.\\
9: Classified as O7V: by Vijapurkar \&
Drilling~(\cite{vijapurkar93}).\\
10: Member of cluster \object{NGC~6910}\\
11: In DR~21.\\
12: Classified as B0.5V by Hoag \& Applequist~(\cite{hoag65})\\
13: In DR~22, member of cluster 11 of Le Duigou \&
Kn\"odlseder~(\cite{leduigou02}).\\
14: In DR~17, member of cluster 12 of Le Duigou \&
Kn\"odlseder~(\cite{leduigou02}).\\
15: In cluster associated to \object{DR 23} (Bica et
al.~\cite{bica03}).\\
16: Possibly associated to DR~5.\\
17: In DR~22. \\
18: Possibly associated to DR~5.\\
19: Possibly associated to the radiosource \object{MITGJ2029+3936}
(Griffith et al.~\cite{griffith91}).\\
20: Possibly associated to DR~23.\\
21: In DR~17.\\
22: In DR~17.\\
23: Possibly associated to DR~5.\\
24: Probably associated to \object{IRAS~20346+3926A}.\\
25: In cluster NGC~6910 (Delgado \& Alfaro~\cite{delgado00}).\\
26: In DR~21.\\
27: Possibly associated to DR~22.\\
}
% End \longtab

\end{document}